\documentclass[iop]{emulateapj}
\usepackage{apjfonts}
\usepackage{graphicx}
\usepackage{psfig}
\usepackage{natbib}
\slugcomment{{\sc Accepted to ApJ}}
\makeatletter
\newcommand{\oii}{[O~{\sc ii}]~$\lambda$3727}
\newcommand{\oiib}{[O~{\sc ii}]~$\lambda$3726}
\newcommand{\oiir}{[O~{\sc ii}]~$\lambda$3729}
\newcommand{\neiii}{[Ne~{\sc iii}]~$\lambda$3868}
\newcommand{\siia}{[S~{\sc ii}]~$\lambda$4069}
\newcommand{\ciiorl}{C~{\sc ii}~$\lambda$4267}
\newcommand{\oiiia}{[O~{\sc iii}]~$\lambda$4363}
\newcommand{\oiiorl}{O~{\sc ii}~$\lambda$4649}
\newcommand{\heii}{He~{\sc ii}~$\lambda$4686}
\newcommand{\ariv}{[Ar~{\sc iv}]~$\lambda$4740}
\newcommand{\oiii}{[O~{\sc iii}] $\lambda\lambda$4959,~5007}
\newcommand{\oiiib}{[O~{\sc iii}]~$\lambda$4959}
\newcommand{\oiiir}{[O~{\sc iii}]~$\lambda$5007}
\newcommand{\niia}{[N~{\sc ii}]~$\lambda$5755}
\newcommand{\siiia}{[S~{\sc iii}]~$\lambda$6312}
\newcommand{\nii}{[N~{\sc ii}] $\lambda\lambda$6548,~6583}
\newcommand{\niib}{[N~{\sc ii}]~$\lambda$6548}
\newcommand{\niir}{[N~{\sc ii}]~$\lambda$6583}
\newcommand{\sii}{[S~{\sc ii}] $\lambda \lambda$6716,~6731}
\newcommand{\siib}{[S~{\sc ii}]~$\lambda$6716}
\newcommand{\siir}{[S~{\sc ii}]~$\lambda$6731}
\newcommand{\ariii}{[Ar~{\sc iii}]~$\lambda$7135}
\newcommand{\oiia}{[O~{\sc ii}] $\lambda\lambda$7320,~7330}
\newcommand{\oiiab}{[O~{\sc ii}]~$\lambda$7320}
\newcommand{\oiiar}{[O~{\sc ii}]~$\lambda$7330}

\newcommand{\halpha}{H$\alpha$}
\newcommand{\hbeta}{H$\beta$}
\newcommand{\hgamma}{H$\gamma$}

\newcommand{\oiis}{[O~{\sc ii}]}
\newcommand{\oiiis}{[O~{\sc iii}]}
\newcommand{\niis}{[N~{\sc ii}]}
\newcommand{\niiis}{[N~{\sc iii}]}
\newcommand{\siis}{[S~{\sc ii}]}

\newcommand{\te}{$T_{\rm e}$}
\newcommand{\tii}{$T_2$}
\newcommand{\tiii}{$T_3$}
\newcommand{\teoii}{$T_{\rm e}$[O~{\sc ii}]}
\newcommand{\teoiii}{$T_{\rm e}$[O~{\sc iii}]}
\newcommand{\tenii}{$T_{\rm e}$[N~{\sc ii}]}
\newcommand{\tesii}{$T_{\rm e}$[S~{\sc ii}]}

\newcommand{\msun}{M$_\odot$}
\newcommand{\mstar}{$M_\star$}
\newcommand{\rtt}{$R_{23}$}
\newcommand{\hii}{H~{\sc ii}}

\makeatother
\begin{document}

\title{The Mass--Metallicity Relation with the Direct Method on Stacked Spectra
  of SDSS Galaxies} \author{Brett H.~Andrews\altaffilmark{1} \& Paul
  Martini\altaffilmark{1}} \altaffiltext{1}{Department of Astronomy, The Ohio
  State University, 140 West 18th Avenue, Columbus, OH 43210,
  andrews@astronomy.ohio-state.edu}

\begin{abstract}
The relation between galaxy stellar mass and gas-phase metallicity is a
sensitive diagnostic of the main processes that drive galaxy evolution, namely
cosmological gas inflow, metal production in stars, and gas outflow via galactic
winds. We employed the direct method to measure the metallicities of
$\sim$200,000 star-forming galaxies from the SDSS that were stacked in bins of
(1) stellar mass and (2) both stellar mass and star formation rate (SFR) to
significantly enhance the signal-to-noise ratio of the weak \oiiia\ and \oiia\
auroral lines required to apply the direct method. These metallicity
measurements span three decades in stellar mass from log(\mstar/\msun)~
=~7.4--10.5, which allows the direct method mass--metallicity relation to
simultaneously capture the high-mass turnover and extend a full decade lower in
mass than previous studies that employed more uncertain strong line methods. The
direct method mass--metallicity relation rises steeply at low mass
(O/H~$\propto$~\mstar$^{1/2}$) until it turns over at log(\mstar/\msun)~=~8.9
and asymptotes to 12 + log(O/H)~=~8.8 at high mass. The direct method
mass--metallicity relation has a steeper slope, a lower turnover mass, and a
factor of two to three greater dependence on SFR than strong line
mass--metallicity relations. Furthermore, the SFR-dependence appears monotonic
with stellar mass, unlike strong line mass--metallicity relations. We also
measure the N/O abundance ratio, an important tracer of star formation history,
and find the clear signature of primary and secondary nitrogen enrichment.  N/O
correlates tightly with oxygen abundance, and even more so with stellar mass.
\end{abstract}


\defcitealias{tremonti2004}{T04}
\defcitealias{mcgaugh1991}{M91}
\defcitealias{zaritsky1994}{Z94}
\defcitealias{kobulnicky2004}{KK04}
\defcitealias{pilyugin2005}{PT05}
\defcitealias{denicolo2002}{D02}
\defcitealias{kewley2002}{KD02}
\defcitealias{pettini2004}{PP04}

\section{Introduction}
Galaxy metallicities are one of the fundamental observational quantities that
provide information about their evolution.  The metal content of a galaxy is
governed by a complex interplay between cosmological gas inflow, metal
production by stars, and gas outflow via galactic winds.  Inflows dilute the
metallicity of a galaxy in the short term but provide the raw fuel for star
formation on longer timescales.  This gas turns into stars, which convert
hydrogen and helium into heavier elements.  The newly formed massive stars
inject energy and momentum into the gas, driving large-scale outflows that
transport gas and metals out of the galaxy.  The ejected metals can escape the
gravitational potential well of the galaxy to enrich the intergalactic medium or
reaccrete onto the galaxy and enrich the inflowing gas.  This cycling of baryons
in and out of galaxies directly impacts the stellar mass (\mstar), metallicity
($Z$), and star formation rate (SFR) of the galaxies.  Thus, the galaxy stellar
mass--metallicity relation (MZR) and the stellar mass--metallicity--SFR relation
serve as crucial observational constraints for galaxy evolution models that
attempt to understand the build up of galaxies across cosmic time.  Here we
present new measurements of the MZR and the \mstar--$Z$--SFR relation that span
three orders of magnitude in stellar mass with metallicities measured with the
direct method.

The first indication of a correlation between mass and metallicity came when
\citet{lequeux1979} demonstrated the existence of a relation between total mass
and metallicity for irregular and blue compact galaxies.  Subsequent studies
showed that metallicity also correlates with other galaxy properties, such as
luminosity \citep{rubin1984} and rotation velocity \citep{zaritsky1994,
garnett2002}.  The advent of reliable stellar population synthesis models
\citep{bruzual2003} enabled more accurate stellar mass measurements from
spectral energy distributions.  \citet[hereafter T04]{tremonti2004} showed the
existence of a tight correlation between galaxy stellar mass and metallicity
among $\sim$53,000 galaxies from the Sloan Digital Sky Survey
\citep[SDSS;][]{york2000} DR2 \citep{abazajian2003} based on the tellar mass
measurements from \citet{kauffmann2003a}.  The \citetalias{tremonti2004} MZR
increases as roughly O/H~$\propto$~\mstar$^{1/3}$ from
\mstar~=~10$^{8.5}$--10$^{10.5}$~\msun\ and then flattens above
\mstar~$\sim$10$^{10.5}$~\msun.  They found that the scatter in the MZR was
smaller than the scatter in the luminosity--metallicity relation and concluded
that the MZR was more physically motivated.  \citet{lee2006} extended the MZR
down another $\sim$2.5 dex in stellar mass with a sample of local dwarf
irregular galaxies.  The scatter and slope of the \citet{lee2006} MZR are
consistent with the \citetalias{tremonti2004} MZR \citep[cf.,][]{zahid2012a},
but the \citet{lee2006} MZR is offset to lower metallicities by 0.2--0.3 dex.
This offset is likely because \citetalias{tremonti2004} and \citet{lee2006} use
different methods to estimate metallicity.  Later work by \citet{ellison2008}
discovered that galaxies with high SFRs (and larger half-light radii) are
systematically offset to lower metallicities than more weakly star-forming
galaxies at the same stellar mass.  \citet{mannucci2010} and
\citet{laralopez2010} studied this effect in a systematic fashion and
demonstrated that the scatter in the MZR is reduced further by accounting for
SFR.  \citet{mannucci2010} introduced the concept of the fundamental metallicity
relation (FMR) by parametrizing the second-order dependence of the MZR on SFR
with a new abscissa,
\begin{equation}
\mu_\alpha \equiv \mathrm{log}(M_\star) - \alpha \mathrm{log(SFR)},
\label{eqn:fmr}
\end{equation}
where the coefficient $\alpha$ is chosen to minimize the scatter in the
relation.  We will refer to this particular parametrization as the FMR but the
general relation as the \mstar--$Z$--SFR relation.  Interestingly,
\citet{mannucci2010} and \citet{laralopez2010} found that the \mstar--$Z$--SFR
relation does not evolve with redshift up to $z$~$\sim$~2.5, as opposed to the
MZR \citep{erb2006, maiolino2008, zahid2011a, moustakas2011}.  However, this
result depends on challenging high redshift metallicity measurements,
specifically the \citet{erb2006} sample of stacked galaxy spectra at $z \sim
2.2$ and the \citet{maiolino2008} sample of nine galaxies at $z \sim 3.5$.

Galaxy evolution models aim to reproduce various features of the MZR and
\mstar--$Z$--SFR relation, specifically their slope, shape, scatter, and
evolution.  The most distinguishing characteristic of the shape of the MZR is
that it appears to flatten and become independent of mass at
\mstar~$\sim$~10$^{10.5}$~\msun.  The canonical explanation is that this
turnover reflects the efficiency of metal ejection from galaxies because the
gravitational potential wells of galaxies at and above this mass scale are too
deep for supernova-driven winds to escape \citep{dekel1986, dekel2003,
tremonti2004}.  In this scenario, the metallicity of these galaxies approaches
the effective yield of the stellar population.  However, recent simulations by
\citet{oppenheimer2006}, \citet{finlator2008}, and \citet{dave2011b, dave2011a}
show that winds characterized by a constant velocity and constant mass-loading
parameter (mass outflow rate divided by SFR; their \textit{cw} simulations),
which were intended to represent supernova-driven winds, result in a MZR that
fails to qualitatively match observations.  The \textit{cw} simulations produce
a MZR that is flat with a very large scatter at low mass, yet becomes steep
above the blowout mass, which is the critical scale above which all metals are
retained.  Instead, they find that their simulations with momentum-driven winds
\citep{murray2005, zhang2012} best reproduce the slope, shape, scatter, and
evolution of the MZR because the wind velocity scales with the escape velocity
of the halo.  Their model naturally produces a FMR that shows little evolution
since $z$~=~3, consistent with observations \citep{mannucci2010, richard2011,
cresci2012}.  However, their FMR does not quite reach the low observed scatter
reported by \citet{mannucci2010}.  Additionally, they find that the coefficient
relating \mstar\ and SFR that minimizes the scatter in the FMR is different
from the one found by \citet{mannucci2010}.  While there is hardly a consensus
among galaxy evolution models about how to produce the MZR and \mstar--$Z$--SFR
relation, it is clear that additional observational constraints would improve
the situation.  So far, the overall normalization of the MZR and the
\mstar--$Z$--SFR relation have been mostly ignored by galaxy evolution models
due to uncertainties in the nucleosynthetic yields used by the models and the
large (up to a factor of five) uncertainties in the normalization of the
observed relations caused by systematic offsets among metallicity calibrations.
If these uncertainties could be reduced, then the normalization could be used
as an additional constraint on galaxy evolution models.

The current metallicity and the metal enrichment history also have implications
for certain types of stellar explosions.  There is mounting evidence that long
duration gamma ray bursts \citep{stanek2006}, over-luminous type~II supernovae
\citep{stoll2011}, and super-Chandrasekhar type~Ia supernovae \citep{khan2011}
preferentially occur in low metallicity environments.  The progenitors of long
gamma ray bursts and over-luminous type II supernovae are thought to be massive
stars and the nature of their explosive death could plausibly depend on their
metallicity.  The cause of the association between super-Chandrasekhar type Ia
supernovae and low metallicity environments is still highly uncertain because
the progenitors are not well known.  Nevertheless, accurate absolute
metallicities for the host galaxies of the progenitors of gamma ray bursts,
over-luminous supernovae, and super-Chandrasekhar type Ia supernovae will help
inform the models of stellar evolution and explosions that attempt to explain
these phenomena.

The uncertainty in the absolute metallicity scale can be traced to differences
between the two main methods of measuring metallicity: the direct method and
strong line method.  The direct method utilizes the flux ratio of auroral to
strong lines to measure the electron temperature of the gas, which is a good
proxy for metallicity because metals are the primary coolants of \hii\ regions.
This flux ratio is sensitive to temperature because the auroral and strong lines
originate from the second and first excited states, respectively, and the
relative level populations depend heavily on electron temperature.  The electron
temperature is a strong function of metallicity, such that hotter electron
temperatures correspond to lower metallicities.  In the direct method, the
electron temperature estimate is the critical step because the uncertainty in
metallicity is nearly always dominated by the uncertainty in the electron
temperature.  The strong line method uses the flux ratios of the strong lines,
which do not directly measure the metallicity of the \hii\ regions but are
metallicity-sensitive and can be calibrated to give approximate metallicities.
The direct method is chosen over strong line methods when the auroral lines can
be detected, but these lines are often too weak to detect at high metallicity.
The strong lines, on the other hand, are much more easily detected than the
auroral lines, particularly in metal-rich objects.  Consequently, the strong
line method can be used across a wide range of metallicity and on much lower
signal-to-noise ratio (SNR) data, so nearly all metallicity studies of large
galaxy samples employ the strong line method.  Despite the convenience of the
strong line method, the relationship between strong line ratios and metallicity
is complicated due to the sensitivity of the strong lines to the hardness of the
incident stellar radiation field and the excitation and ionization states of the
gas.  Thus, strong line ratios must be calibrated (1) empirically with direct
method metallicities, (2) theoretically with photoionization models, or (3)
semi-empirically with a combination of direct method metallicities and
theoretically calibrated metallicities.  Unfortunately, the three classes of
calibrations do not generically produce consistent metallicities.  For example,
metallicities determined with theoretical strong line calibrations are
systematically higher than those from the direct method or empirical strong line
calibrations by up to $\sim$0.7 dex \citep[for a detailed discussion
see][]{moustakas2010, stasinska2010}.  The various strong line methods also
exhibit systematic disagreements as a function of metallicity and perform better
or poorer in certain metallicity ranges.

The cause of the discrepancy between direct method metallicities and
theoretically calibrated metallicities is currently unknown.  As recognized by
\citet{peimbert1967}, the electron temperatures determined in the direct method
might be overestimated in the presence of temperature gradients and/or
fluctuations in \hii\ regions.  Such an effect would cause the direct method
metallicities to be biased low \citep{stasinska2005, bresolin2008}.  A similar
result could arise if the traditionally adopted electron energy distribution is
different from the true distribution, as suggested by \citet{nicholls2012}.
Alternatively, the photoionization models that serve as the basis for the
theoretical strong line calibrations, such as {\sc cloudy} \citep{ferland1998}
and {\sc mappings} \citep{sutherland1993}, make simplifying assumptions in their
treatment of \hii\ regions that may result in overestimated metallicities, such
as the geometry of the nebula or the age of the ionizing stars \citep[see][for a
thorough discussion of these issues]{moustakas2010}; however, no one particular
assumption has been conclusively identified to be the root cause of the
metallicity discrepancy.

In this work, we address the uncertainty in the absolute metallicity scale by
using the direct method on a large sample of galaxies that span a wide range of
metallicity.  The uniform application of the direct method also provides more
consistent metallicity estimates over a broad range in stellar mass.  While the
auroral lines used in the direct method are undetected in most galaxies, we have
stacked the spectra of many galaxies (typically hundreds to thousands) to
significantly enhance the SNR of these lines.  In Section \ref{sec:method}, we
describe the sample selection, stacking procedure, and stellar continuum
subtraction.  Section \ref{sec:metal_cal} describes the direct method and strong
line metallicity calibrations that we use.  In Section \ref{sec:gal_comp}, we
demonstrate that mean galaxy properties can be recovered from stacked spectra.
We show the electron temperature relations for the stacks in Section
\ref{sec:te} and argue that \teoii\ is a better tracer of oxygen abundance than
\teoiii\ in Section \ref{sec:abund}.  Section \ref{sec:mzr_fmr} shows the main
results of this study: the MZR and \mstar--$Z$--SFR relation with the direct
method.  In Section \ref{sec:no}, we present the direct method N/O relative
abundance as a function of O/H and stellar mass.  Section \ref{sec:discussion}
details the major uncertainties in metallicity measurements and the implications
for the physical processes that govern the MZR and \mstar--$Z$--SFR relation.
Finally, we present a summary of our results in Section \ref{sec:summary}.  For
the purpose of discussing metallicities relative to the solar value, we adopt
the solar oxygen abundance of 12~+~log(O/H)~=~8.86 from \citet{delahaye2006}.
Throughout this work, stellar masses and SFRs are in units of \msun\ and
\msun~yr$^{-1}$, respectively.  We assume a standard $\Lambda$CDM cosmology with
$\Omega_\mathrm{m} = 0.3$, $\Omega_\Lambda = 0.7$, and $H_0 =
70$~km~s$^{-1}$~Mpc$^{-1}$.


\section{Method}
\label{sec:method}
\subsection{Sample Selection}
The observations for our galaxy sample come from the SDSS Data Release 7
\citep[DR7;][]{abazajian2009}, a survey that includes $\sim$930,000 galaxies
\citep{strauss2002} in an area of 8423 square degrees.  The parent sample for
this study comes from the MPA-JHU catalog\footnote{Available at
http://www.mpa-garching.mpg.de/SDSS/DR7/} of 818,333 unique galaxies which have
derived stellar masses \citep{kauffmann2003a}, SFRs \citep{brinchmann2004,
salim2007}, and metallicities \citepalias{tremonti2004}.  We chose only
galaxies with reliable redshifts ($\sigma_{z}$~<~0.001) in the range
0.027~<~$z$~<~0.25 to ensure that the \oii\ line and the \oiia\ lines fall
within the wavelength range of the SDSS spectrograph (3800--9200~\AA).

We discard galaxies classified as AGN because AGN emission line ratios may
produce spurious metallicity measurements.  We adopt the \citet{kauffmann2003b}
criteria (their Equation 1) to differentiate between star-forming galaxies and
AGN, which employs the emission line ratios that define the Baldwin, Phillips,
and Terlevich (1981) (BPT) diagram:
\begin{eqnarray}
\lefteqn{\!\!\!\!\!\!\!\!\!\!\!\!\!\!\!\!\!\!\!\!\!\!\!\!{\rm log} ([{\rm O} \;
    \textsc{iii}] \, \lambda 5007 / {\rm H} \beta) >} \nonumber \\ & & 0.61
    [{\rm log} ({[{\rm N} \; \textsc{ii}] \, \lambda 6583 / {\rm H}\alpha}) -
    0.05]^{-1} + 1.3.
\label{eqn:bpt}
\end{eqnarray}
We follow the \citetalias{tremonti2004} SNR thresholds for emission lines.
Specifically, we restrict our sample to galaxies with \hbeta, \halpha, and
\niir\ detected at >5$\sigma$.  Further, we apply the AGN--star-forming galaxy
cut (Equation \ref{eqn:bpt}) to galaxies with >3$\sigma$ detections of \oiiir.
We also select galaxies with \oiiir~<~3$\sigma$ but log(\niir/\halpha)~<~$-$0.4
as star-forming to include high metallicity galaxies with weak \oiiir.

At the lowest stellar masses (log[\mstar]~<~8.6), this initial sample is
significantly contaminated by spurious galaxies, which are actually the
outskirts of more massive galaxies and were targeted due to poor photometric
deblending.  We remove galaxies whose photometric flags include {\sc
deblend\_nopeak} or {\sc deblended\_at\_edge}.  We also visually inspected all
galaxies with log(\mstar)~<~8.6 and discarded any that suffered from obvious
errors in the stellar mass determination (again, likely as a result of
off-center targeting of a much more massive galaxy).

After all of our cuts, the total number of galaxies in our sample is 208,529 and
the median redshift is $z$~=~0.078.  At this redshift, the 3'' diameter SDSS
aperture will capture light from the inner 2.21~kpc of a galaxy.  Since the
central regions of galaxies will tend to be more metal-rich \citep{searle1971},
the metallicities measured from these observations will likely be biased high
due to the aperture size relative to angular extent of the galaxies.  However,
we expect this bias is small for most galaxies \citep[for a more detailed
discussion see][]{tremonti2004, kewley2005}. In particular, the galaxies with
very low stellar masses and metallicities that define the low mass end of the
MZR tend to be compact and have homogeneous metallicities \citep[e.g.,
][]{kobulnicky1997b}, although many of these are excluded by the criteria
proposed by \citet{kewley2005}.

\subsection{Stacking Procedure}
\label{sec:stacking}

\begin{figure*}
\centerline{\includegraphics[width=19cm]{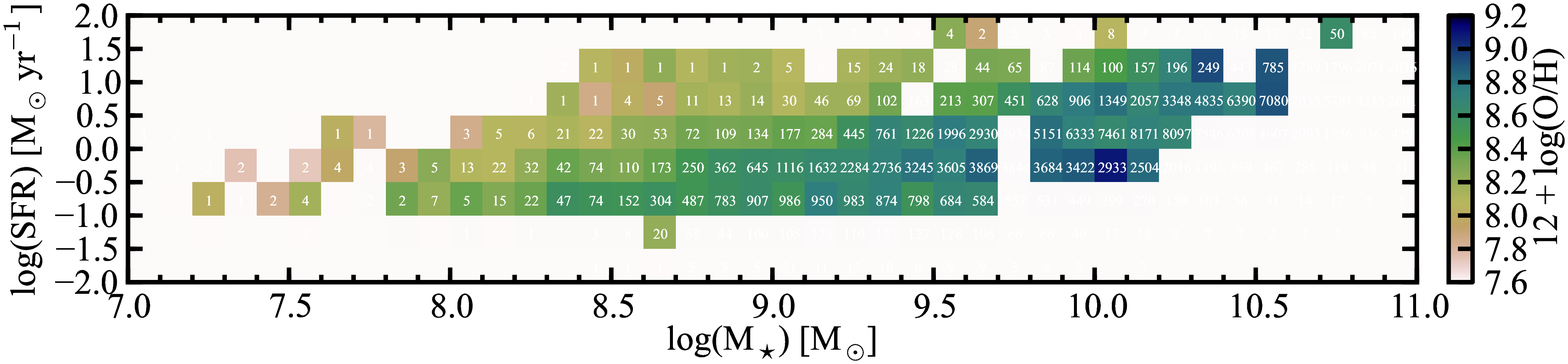}}
\caption{Number of galaxies and direct method metallicity as a function of
  \mstar\ and SFR.  The squares represent each \mstar--SFR stack, the number of
  galaxies is indicated by the white text, and the color scale corresponds to
  the metallicity.  For reference, the \citet{tremonti2004} MZR covers
  log(\mstar)~=~8.5--11.5, and the \citet{mannucci2010} FMR spans
  log(\mstar)~=~9.1--11.35 and log(SFR)~=~$-$1.45$\rightarrow$0.80.
\label{fig:msfr}}
\end{figure*}
%
\begin{figure*}
\centerline{\includegraphics[width=19cm]{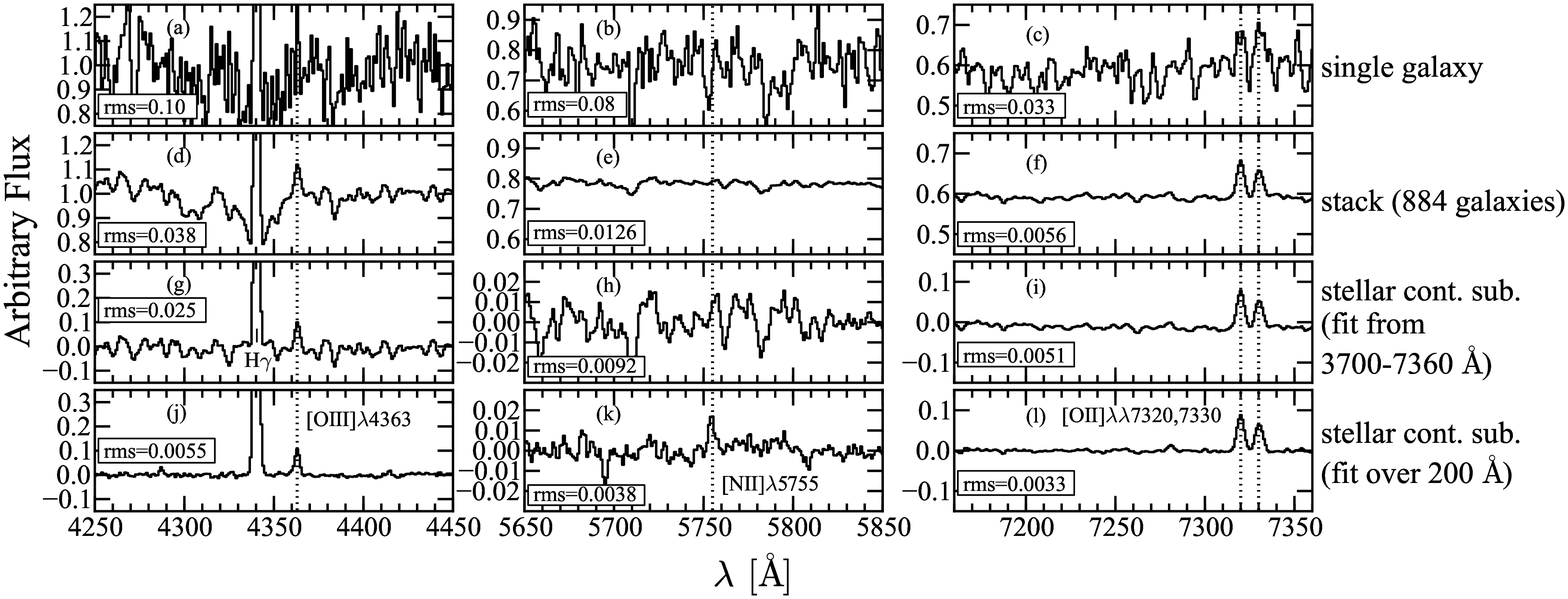}}
\caption{Sample spectra from the log(\mstar)~=~8.7--8.8 ($N_\mathrm{gal}$~=~884)
  stack.  From left to right, the three columns show the \oiiia, \niia, and
  \oiia\ auroral lines.  From top to bottom, the four rows correspond to the
  reduced spectrum of a single galaxy, the spectrum of the stack, the spectrum
  of the stack after the removal of the stellar continuum (fit from
  3700--7360~\AA), and the spectrum of the stack after the removal of the
  stellar continuum (fit to a 200~\AA\ window near the emission line of
  interest).  The continuum rms of each spectrum near the relevant emission line
  is given in the inset of each panel.
\label{fig:sample_spec}}
\end{figure*}

The primary motivation for this investigation is to measure the metallicity of
galaxies with the direct method.  The main challenge is that the weak \oiiia\
and \oiia\ auroral lines are undetected in most of the individual spectra.  To
improve the SNR of the spectra, we stacked galaxies that are expected to have
similar metallicities and hence line ratios.  Given the tightness of the MZR
and \mstar--$Z$--SFR relation, it is reasonable to expect that galaxies at a
given stellar mass, or simultaneously a given stellar mass and SFR, will have
approximately the same metallicity.  Thus, we have created two sets of galaxy
stacks: (1) galaxies binned in 0.1 dex in \mstar\ from
log(\mstar/\msun)~=~7.0~to~11.0 (hereafter \mstar\ stacks) and (2) galaxies
binned in 0.1 dex in \mstar\ from log(\mstar/\msun)~=~7.0~to~11.0 {\it and} 0.5
dex in SFR from log(SFR/[\msun~ yr$^{-1}$])~=~$-$2.0~to~2.0 (hereafter
\mstar--SFR stacks).  We adopt the total stellar mass \citep{kauffmann2003a}
and the total SFR \citep{brinchmann2004, salim2007} values from the MPA-JHU
catalog, as opposed to these quantities calculated only for the light within
the fiber.  For convenience, we will refer to the stacks by the type of stack
with a subscript and a superscript to denote the upper and lower bounds of
log(\mstar) or log(SFR) (e.g., \mstar$_{8.7}^{8.8}$ is the \mstar\ stack with
log[\mstar/\msun]~=~8.7--8.8, and SFR$_{0.0}^{0.5}$ corresponds to the
\mstar--SFR stacks with log[SFR/{\msun~yr$^{-1}$}]~=~0.0--0.5).  Figure
\ref{fig:msfr} shows the number of galaxies in each \mstar--SFR stack (each box
represents a stack) with a measured metallicity (indicated by the color
coding).

We stacked galaxy spectra that have been processed with the SDSS reduction
pipeline \citep{stoughton2002}.  First, we corrected for Milky Way reddening
with the extinction values from \citet{schlegel1998}.  Then, the individual
galaxy spectra were shifted to the rest frame with the redshifts from the
MPA/JHU catalog.  Next, we linearly interpolated the spectra onto a universal
grid (3700--7360~\AA; $\Delta\lambda$~=~1~\AA) in linear-$\lambda$ space.  This
interpolation scheme conserves flux in part because the wavelength spacing of
the grid is narrower than the width of bright emission lines.  The spectra were
then normalized to the mean flux from 4400--4450~\AA.  Finally, the spectra
were co-added (i.e., we took the mean flux in each wavelength bin) to form the
stacked spectra (see Section \ref{sec:gal_comp} for comparisons between the
electron temperatures and metallicities of stacks and individual galaxies).

Figure \ref{fig:sample_spec} shows the SNR increase of the \oiiia\ (left
column), \niia\ (middle column), and \oiia\ (right column) lines as the spectra
are processed from a typical single galaxy spectrum (top row) to the stacked
spectrum (second row) to the stellar continuum subtracted spectrum (third row;
see Section \ref{sec:scs}) or the narrow wavelength window stellar continuum
subtracted spectrum (bottom row; see Section \ref{sec:scs}).  The spectra in the
top row are from a typical galaxy in the log(\mstar)~=~8.7--8.8 bin; the bottom
three rows show the stacked spectra from the same bin.  In each panel, we report
the continuum root mean square (rms).  The decrease in the continuum noise when
comparing the spectra in the top row to the second row of Figure
\ref{fig:sample_spec} is dramatic.  Further significant noise reduction can be
achieved by removing the stellar continuum (shown in the bottom two rows of
Figure \ref{fig:sample_spec}), as we describe in Section \ref{sec:scs}.

\subsection{Stellar Continuum Subtraction}
\label{sec:scs}

\begin{deluxetable}{ccc}
\tabletypesize{\small}
\tablecaption{Wavelength Fit and Mask Ranges of Measured Lines
\label{table:wave}}
\tablewidth{0pt}
\tablehead{
\colhead{Line} &
\colhead{Fit Range} &
\colhead{Mask Range} \\
\colhead{} &
\colhead{[\AA]} &
\colhead{[\AA]} \\
\colhead{(1)} &
\colhead{(2)} &
\colhead{(3)}
}
\startdata
{}[O~{\sc ii}]~$\lambda$3727  &  3700--4300  &  3710--3744 \\
{}[Ne~{\sc iii}]~$\lambda$3868  &  3800--4100  &  3863--3873 \\
{}[S~{\sc ii}]~$\lambda$4069  &  3950--4150  &  \nodata \\
{}H$\gamma$~$\lambda$4340  &  4250--4450  &  4336--4344 \\
{}[O~{\sc iii}]~$\lambda$4363  &  4250--4450  &  4360--4366 \\
{}He~{\sc ii}~$\lambda$4686  &  4600--4800  &  4680--4692 \\
{}[Ar~{\sc iv}]~$\lambda$4740  &  3700--7360  &  \nodata \\
{}H$\beta$~$\lambda$4861  &  3700--7360  &  4857--4870 \\
{}[O~{\sc iii}]~$\lambda$4959  &  3700--7360  &  4954--4964 \\
{}[O~{\sc iii}]~$\lambda$5007  &  3700--7360  &  5001--5013 \\
{}[N~{\sc ii}]~$\lambda$5755  &  5650--5850  &  5753--5757 \\
{}[S~{\sc iii}]~$\lambda$6312  &  6100--6500  &  6265--6322 \\
{}[N~{\sc ii}]~$\lambda$6548  &  3700--7360  &  6528--6608 \\
{}H$\alpha$~$\lambda$6563  &  3700--7360  &  6528--6608 \\
{}[N~{\sc ii}]~$\lambda$6583  &  3700--7360  &  6528--6608 \\
{}[S~{\sc ii}]~$\lambda$6716  &  3700--7360  &  6696--6752 \\
{}[S~{\sc ii}]~$\lambda$6731  &  3700--7360  &  6696--6752 \\
{}[Ar~{\sc iii}]~$\lambda$7135  &  7035--7235  &  7130--7140 \\
{}[O~{\sc ii}]~$\lambda$7320  &  7160--7360  &  7318--7322 \\
{}[O~{\sc ii}]~$\lambda$7330  &  7160--7360  &  7328--7332 \\
\enddata
\tablecomments{Column (1): Emission lines.  Column (2): The wavelength range
  of the stellar continuum fit.  Column (3) The wavelength range of the stellar
  continuum fit that was masked out.}
\end{deluxetable}

Stacking the spectra increases the SNR, but it is important to fit and subtract
the stellar continuum to detect and accurately measure the flux of these lines,
especially \oiiia\ due to its proximity to the \hgamma\ stellar absorption
feature.  We subtracted the stellar continuum with synthetic template galaxy
spectra created with the {\sc starlight} stellar synthesis code
\citep{cidfernandes2005}, adopted the \citet{cardelli1989} extinction law, and
masked out the locations of the emission lines.  The synthetic spectra were
created from a library of 300 empirical {\sc miles} spectral templates
\citep[data as obtained from the {\sc miles}
website\footnote{http://miles.iac.es/}]{sanchezblazquez2006, cenarro2007,
vazdekis2010, falconbarroso2011}.  The {\sc miles} templates provided an
excellent fit to the stellar continuum (see bottom two rows of Figure
\ref{fig:sample_spec}).  We note the {\sc miles} templates yielded better fits
to the very high SNR spectra than the \citet{bruzual2003} spectral templates,
based on the {\sc stelib} \citep{leborgne2003} library.

We performed stellar template fits to the entire spectral range, select
subregions centered on weak lines of interest, and subregions around the strong
lines blueward of 4000~\AA.  The latter are situated among a forest of stellar
absorption lines.  The line fluxes of the strong emission lines redward of
4000~\AA\ (\hbeta, \oiii, \halpha, \nii, and \sii) were measured from the
spectrum where the stellar continuum was fit over the full wavelength range of
our stacked spectra ($\lambda$~=~3600--7360~\AA; see third row of Figure
\ref{fig:sample_spec}).  The stellar continuum subtraction near weak emission
lines (\siia, \oiiia, \heii, \niia, \siiia, \ariv, and \oiia) and blue strong
emission lines (\oii\ and \neiii) was improved if the stellar continuum fit was
restricted to limited wavelength ranges within a few 100~\AA\ of the line of
interest (compare the third and bottom rows of Figure \ref{fig:sample_spec}).
For the weak lines and blue strong lines, we measured the line fluxes from the
stellar continuum subtracted spectra within these narrow wavelength windows
(details are listed in Table \ref{table:wave}).  In order to compare the line
fluxes across regions with different stellar continuum subtraction (e.g., from
portions of the spectrum that were fit with smaller wavelength ranges), we
denormalized the spectra after the {\sc starlight} fit.

\subsection{Automated Line Flux Measurements}

\begin{deluxetable}{cll}
\tabletypesize{\small}
\tablecaption{Line Fluxes \label{table:flux}}
\tablewidth{0pt}
\tablehead{
\colhead{Column} &
\colhead{Format} &
\colhead{Description}
}
\startdata
1  &   F4.1   &  Lower stellar mass limit of the stack \\
2  &   F4.1   &  Upper stellar mass limit of the stack \\
3  &   F4.1   &  Lower SFR limit of the stack \\
4  &   F4.1   &  Upper SFR limit of the stack \\
5  &   I5     &  Number of galaxies in the stack \\
6  &   F6.3   &  Median stellar mass of the stack \\
7  &   F6.3   &  Median SFR of the stack \\
8  &   F6.2   &  \oii\ line flux \\
9  &   F5.2   &  Error on \oii\ line flux \\
10  &  F5.2   &  \neiii\ line flux \\
11  &  F4.2   &  Error on \neiii\ line flux \\
12  &  F4.2   &  \siia\ line flux \\
13  &  F4.2   &  Error on \siia\ line flux \\
14  &  F6.2   &  \hgamma\ line flux \\
15  &  F5.2   &  Error on \hgamma\ line flux \\
16  &  F5.2   &  \oiiia\ line flux \\
17  &  F4.2   &  Error on \oiiia\ line flux \\
18  &  F4.2   &  \heii\ line flux \\
19  &  F4.2   &  Error on \heii\ line flux \\
20  &  F4.2   &  \ariv\ line flux \\
21  &  F4.2   &  Error on \ariv\ line flux \\
22  &  F6.2   &  \oiiib\ line flux \\
23  &  F4.2   &  Error on \oiiib\ line flux \\
24  &  F6.2   &  \oiiir\ line flux \\
25  &  F5.2   &  Error on \oiiir\ line flux \\
26  &  F4.2   &  \niia\ line flux \\ 
27  &  F4.2   &  Error on \niia\ line flux \\ 
28  &  F4.2   &  \siiia\ line flux \\
29  &  F4.2   &  Error on \siiia\ line flux \\
30  &  F5.2   &  \niib\ line flux \\ 
31  &  F4.2   &  Error on \niib\ line flux \\ 
32  &  F6.2   &  \halpha\ line flux \\
33  &  F5.2   &  Error on \halpha\ line flux \\
34  &  F6.2   &  \niir\ line flux \\
35  &  F4.2   &  Error on \niir\ line flux \\
36  &  F6.2   &  \siib\ line flux \\ 
37  &  F4.2   &  Error on \siib\ line flux \\ 
38  &  F5.2   &  \siir\ line flux \\ 
39  &  F4.2   &  Error on \siir\ line flux \\ 
40  &  F4.2   &  \ariii\ line flux \\
41  &  F4.2   &  Error on \ariii\ line flux \\
42  &  F4.2   &  \oiiab\ line flux \\
43  &  F4.2   &  Error on \oiiab\ line flux \\
44  &  F4.2   &  \oiiar\ line flux \\
45  &  F4.2   &  Error on \oiiar\ line flux \\
\enddata

\tablecomments{This table is published in its entirety in the electronic edition
  of the journal.  The column names are shown here for guidance regarding its
  form and content.}

\end{deluxetable}

We used the {\it specfit} task \citep{kriss1994} in the {\sc iraf/stsdas}
package to automatically fit emission lines with a $\chi^2$ minimization
algorithm.  We simultaneously fit a flat continuum and Gaussian line profiles
for the emission lines, even if lines were blended.  For doublets, we fixed the
width of the weaker line by pinning its velocity width to the stronger line
(\oiib\ to \oiir, \oiiib\ to \oiiir, \niib\ to \niir, \siir\ to \siib, and
\oiiar\ to \oiiab).  We also included the continuum rms of the spectrum as an
input to the fitting procedure.  After experimenting with several different
$\chi^2$ minimization algorithms implemented within {\it specfit}, we chose the
simplex algorithm because of its consistent convergence, particularly for weak
lines.  Line fluxes measured by {\it specfit} generally agreed well with line
fluxes measured interactively with the OSU {\sc liner} package.  The
uncertainty in the line flux is based on the $\chi^2$ fit returned from {\it
specfit}.  Finally, all line fluxes were corrected for reddening with the
extinction law from \citet{cardelli1989} and the assumption that the intrinsic
ratio of the Balmer lines is set by case B recombination
(H$\alpha$/H$\beta$~=~2.86 for \te~=~10,000~K).  We adopted a fixed
H$\alpha$/H$\beta$ ratio, even though it is a weak function of electron
temperature.  For the log(\mstar/\msun)~=~10.0--10.1 stack (\teoii~=~7200~K),
whose oxygen abundance is dominated by O$^+$ (i.e., a stack where the potential
effect would be maximal due to the long wavelength baseline between \oii\ and
\oiia), this effect would decrease log(O$^+$/H$^+$) by $\sim$0.07 dex.  The
line fluxes are presented in an online table whose columns are described in
Table \ref{table:flux}.

We disregarded lines that were poorly fit (negative flux, uncertainty in
central wavelength $>$1~\AA, had uncertainty in the velocity width of
$>$100~km/s, or had low SNR [<5$\sigma$]).  Further care was taken to ensure
the robustness of \oiiia\ flux measurements.  As \mstar\ increased to moderate
values (log[\mstar]~$>$~9.0), an unidentified emission feature at 4359~\AA\
became blended with the \oiiia\ line, which limited the SNR of the line flux
measurement independent of the continuum rms.  We are unsure of the origin of
this feature, but it could be caused by an over-subtraction in the stellar
continuum fit.  We simultaneously fit the 4359~\AA\ feature and \oiiia\ and
pinned the velocity width of both lines to H$\gamma$.  If 4359\AA~>~0.5~\oiiia,
then we determined that \oiiia\ could not be robustly fit.  If \oiiia\ could be
well fit, we refit it with a single Gaussian whose velocity width was pinned to
\hgamma.  The line flux measurements from the single Gaussian fitting agreed
better with interactive line flux measurements than the deblended line flux
measurements.  The remaining weak lines are in regions without strong stellar
absorption features.  Often, the \oiia\ lines could be detected in the stacked
spectra without the stellar continuum fit (see Figure \ref{fig:sample_spec}f).
The \niia\ and \siia\ auroral lines were usually too weak to be detected
without stellar continuum subtraction.

Optical recombination lines, such as \ciiorl\ and \oiiorl, are also sensitive to
metallicity. Unlike auroral lines, they are almost independent of temperature,
so they could provide a useful check on the direct method metallicities.
Unfortunately, optical recombination lines tend to be very weak (e.g., the
median \oiiorl/\oiiia\ ratio of five extragalactic \hii\ regions studied by
\citealt{esteban2009} was 0.08), and we did not detect them in the stacked
spectra.



\section{Electron Temperature and Direct Abundance Determination}
\label{sec:metal_cal}
\subsection{Electron Temperatures}
\label{sec:te}

Different ionic species probe the temperature of different ionization zones of
\hii\ regions \citep[e.g.,][]{stasinska1982, garnett1992}.  In the two-zone
model, the high ionization zone is traced by \oiiis, and the low ionization zone
is traced by \oiis, \niis, and \siis.  \citet{campbell1986} used the
photoionization models of \citet{stasinska1982} to derive a linear relation
between the temperatures in these zones,
\begin{equation}
T_{\rm e}[{\rm O} \; \textsc{ii}] = T_{\rm e}[{\rm N} \; \textsc{ii}] =
T_{\rm e}[{\rm S} \; \textsc{ii}] = 0.7 T_{\rm e}[{\rm O} \; \textsc{iii}] +
3000,
\label{eqn:t2t3}
\end{equation}
where \te\ is in units of K.  Subsequently, we will refer to this relation as
the \tii--\tiii\ relation (see \citealt{pagel1992} and \citealt{izotov2006} for
alternative formulations of the \tii--\tiii\ relation). This relation is
especially useful to infer the abundance of unseen ionization states, a critical
step in measuring the total oxygen abundance.  While convenient, this
theoretical relation may be one of the biggest uncertainties in the direct
method because it is not definitively constrained by observations due to the
large random errors in the flux of \oiia\ \citep[e.g., see][]{kennicutt2003,
pilyugin2006}.  The high SNR of our stacked spectra enables us to measure the
electron temperature of both the high and low ionization zones for many of our
stacks.

We measured the electron temperature of \oiiis, \oiis, \niis, and \siis\ with
the \textit{nebular.temden} routine \citep{shaw1995} in {\sc iraf/stsdas},
which is based on the five level atom program of \citet{derobertis1987}.  This
routine determines the electron temperature from the flux ratio of the auroral
to strong emission line(s) for an assumed electron density.  The diversity of
these temperature diagnostics are valuable cross-checks and provide an
independent check on the applicability of the \tii--\tiii\ relation; however,
for measuring oxygen abundances, we only use \teoiii\ and \teoii.  The electron
density ($n_{\rm e}$) can be measured from the density sensitive \sii\ doublet
\citep[cf.,][]{cai1993}.  For 6/45 of the \mstar\ stacks and 65/228 of the
\mstar--SFR stacks, \siib~/~\siir\ was above the theoretical maximum ratio of
1.43 \citep{osterbrock1989}, which firmly places these galaxies in the low
density regime, and we assume $n_{\rm e}$~=~100~cm$^{-3}$ for our analysis.
\citet{yin2007} found similar inconsistencies between the theoretical maximum
and measured flux ratios for individual galaxies, which suggests that there
might be a real discrepancy between the maximum observed and theoretical values
of \siib~/~\siir.

\begin{deluxetable*}{rrrrrrrrrrr}
\centering
\tabletypesize{\small}
\tablecaption{Electron Temperatures, Metallicity, and N/O Abundance \label{table:metallicity}}
\tablewidth{0pt}
\tablehead{
\multicolumn{2}{c}{log(\mstar)} &
\multicolumn{2}{c}{log(SFR)} &
\colhead{$N_\mathrm{gal}$} &
\colhead{\teoiii} &
\colhead{\teoii} &
\colhead{\tenii} &
\colhead{\tesii} &
\colhead{12 + log(O/H)} &
\colhead{log(N/O)} \\
\multicolumn{2}{c}{[M$_\odot$]} &
\multicolumn{2}{c}{[M$_\odot$ yr$^{-1}$]} &
\colhead{} &
\colhead{[K]} &
\colhead{[K]} &
\colhead{[K]} &
\colhead{[K]} &
\colhead{[dex]} &
\colhead{[dex]} \\
\colhead{(1)} &
\colhead{(2)} &
\colhead{(3)} &
\colhead{(4)} &
\colhead{(5)} &
\colhead{(6)} &
\colhead{(7)} &
\colhead{(8)} &
\colhead{(9)} &
\colhead{(10)} &
\colhead{(11)}
}
\startdata
\multicolumn{11}{c}{\mstar Stacks} \\
\cline{1-11} \\
7.0  &  7.1  &    &    &  1  &  \nodata  &  \nodata  &  \nodata  &  \nodata  &  \nodata  &  \nodata \\ 
7.1  &  7.2  &    &    &  4  &  \nodata  &  \nodata  &  \nodata  &  \nodata  &  \nodata  &  \nodata \\ 
7.2  &  7.3  &    &    &  4  &  14000 $\pm$ 600  &  \nodata  &  \nodata  &  \nodata  &  \nodata  &  \nodata \\ 
7.3  &  7.4  &    &    &  4  &  17500 $\pm$ 200  &  \nodata  &  \nodata  &  \nodata  &  \nodata  &  \nodata \\ 
7.4  &  7.5  &    &    &  2  &  15700 $\pm$ 200  &  12800 $\pm$ 800  &  \nodata  &  \nodata  &  7.82 $\pm$ 0.03  &  \nodata \\ 
$\vdots$    &  $\vdots$  &  $\vdots$  &  $\vdots$  &  $\vdots$  &  $\vdots$  &  $\vdots$  &  $\vdots$  &  $\vdots$  &  $\vdots$  &  $\vdots$\\\cline{1-11}
\multicolumn{11}{c}{\mstar--SFR Stacks} \\
\cline{1-11} \\
7.0  &  7.1  &  0.0  &  0.5  &  1  &  \nodata  &  \nodata  &  \nodata  &  \nodata  &  \nodata  &  \nodata \\ 
7.1  &  7.2  &  -0.5  &  0.0  &  1  &  \nodata  &  \nodata  &  \nodata  &  \nodata  &  \nodata  &  \nodata \\ 
7.1  &  7.2  &  0.0  &  0.5  &  2  &  \nodata  &  \nodata  &  \nodata  &  \nodata  &  \nodata  &  \nodata \\ 
7.2  &  7.3  &  -1.0  &  -0.5  &  1  &  13400 $\pm$ 500  &  11800 $\pm$ 700  &  \nodata  &  \nodata  &  8.04 $\pm$ 0.04  &  \nodata \\ 
7.2  &  7.3  &  -0.5  &  0.0  &  2  &  \nodata  &  \nodata  &  \nodata  &  \nodata  &  \nodata  &  \nodata \\ 
$\vdots$    &  $\vdots$  &  $\vdots$  &  $\vdots$  &  $\vdots$  &  $\vdots$  &
$\vdots$  &  $\vdots$  &  $\vdots$  &  $\vdots$  &  $\vdots$\\
\enddata
\tablecomments{Column (1): Lower stellar mass limit of the stack.  Column (2):
  Upper stellar mass limit of the stack.  Column (3): Lower SFR limit of the
  stack.  Column (4): Upper SFR limit of the stack.  Column (5): Number of
  galaxies in the stack.  Columns (6)--(9): Electron temperatures for \oiiis,
  \oiis, \niis, and \siis.  Column (10): Direct method metallicity.  Column
  (11): N/O abundance.\\ (This table is published in its entirety in the
  electronic edition of the journal.  A portion is shown here for guidance
  regarding its form and content.)}
\end{deluxetable*}

We calculated the electron temperature and density uncertainties by propagating
the line flux uncertainties with Monte Carlo simulations.  For the simulations,
we generated 1,000 realizations of the line fluxes (Gaussian distributed
according to the 1$\sigma$ uncertainty) and processed these realizations through
\textit{nebular.temden}.  The electron temperatures of the stacks are given in
Table \ref{table:metallicity} (full version available online).

\begin{figure*}
\centerline{\includegraphics[width=19cm]{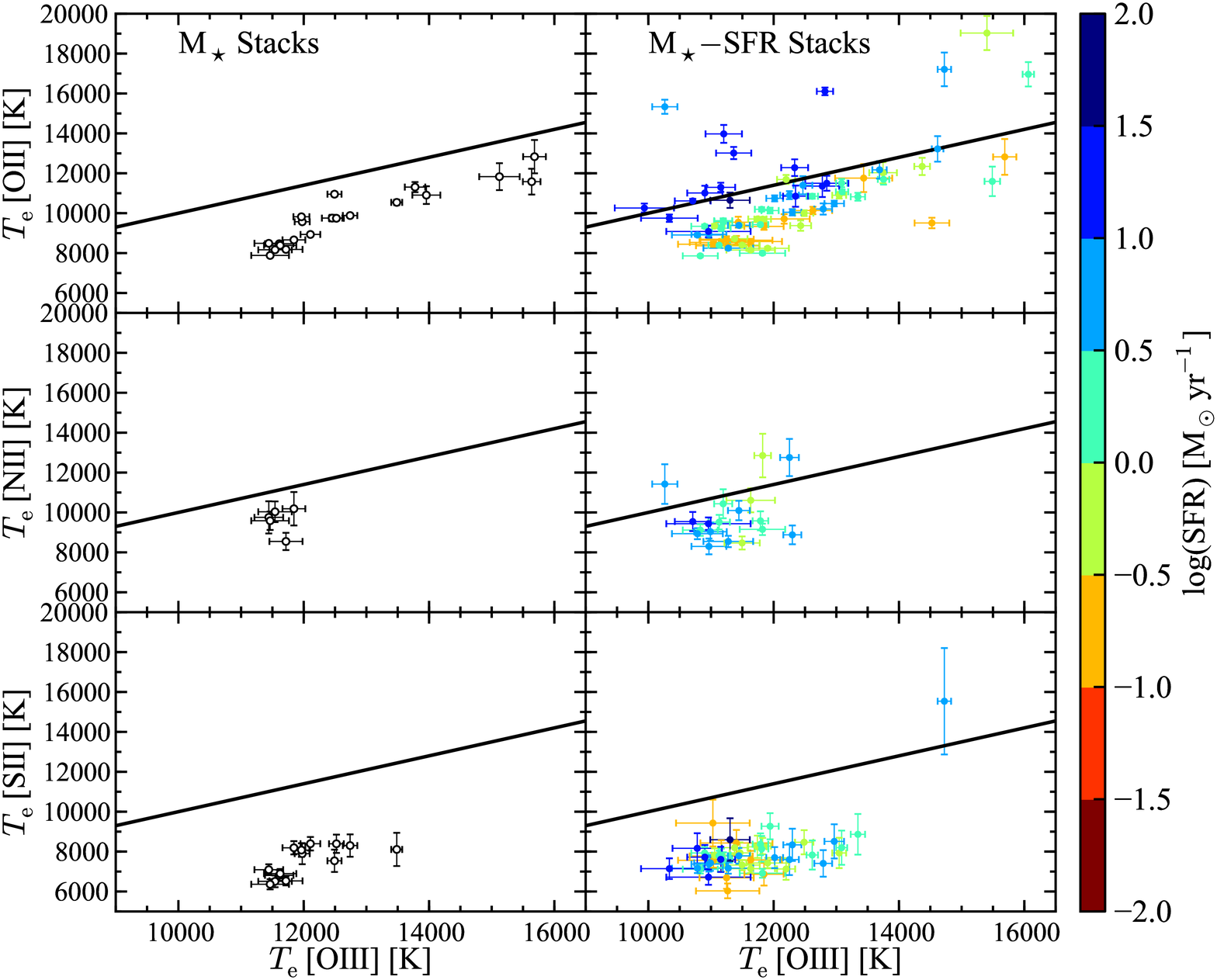}}
\caption{Electron temperatures derived from the \oiis, \niis, and \siis\ line
ratios plotted as a function of electron temperature derived from the \oiiis\
line ratio for the \mstar\ stacks (left column) and \mstar--SFR stacks (right
column; color-coded by SFR).  The lines in the top, middle, and bottom rows show
the \teoii--\teoiii, \tenii--\teoiii, and \tesii--\teoiii\ relations (Equation
\ref{eqn:t2t3}), respectively.  The outlier in the lower right panel is a single
galaxy, so it may not be representative of all galaxies with this stellar mass
and star formation rate.
\label{fig:te_stacks}}
\end{figure*}

In Figure \ref{fig:te_stacks}, we plot the electron temperatures of \oiis,
\niis, and \siis\ against the \oiiis\ electron temperature for the \mstar\
stacks (left column; open circles) and the \mstar--SFR stacks (right column;
circles color-coded by SFR).  For comparison, we show the \tii--\tiii\ relation
(Equation \ref{eqn:t2t3}) as the black line in each panel.  In all three
\te--\te\ plots, the \mstar\ stacks form a tight locus that falls within the
distribution of \mstar--SFR stacks.  The \mstar--SFR stacks show a large
dispersion in \teoii\ at fixed \teoiii\ that is not present in the \mstar\
stacks.  Most of this scatter is due to stacks with SFR$_{1.0}^{1.5}$, which
approach and exceed the \teoii--\teoiii\ relation.  On the other hand, the
\mstar--SFR stacks show little scatter in the \tenii--\teoiii\ and
\tesii--\teoiii\ plots, and they track the \mstar\ stacks in these plots.

The vast majority of the stacks in Figure \ref{fig:te_stacks} fall below the
\tii--\tiii\ relation, independent of the type of stacks (\mstar\ or
\mstar--SFR) or the tracer ion (\oiis, \niis, or \siis).  The multiple
temperature indicators show that the \tii--\tiii\ relation overpredicts the
temperature in the low ionization zone (or underpredicts the temperature in the
high ionization zone).  If we assume that \teoiii\ is accurate (i.e., the
temperature in the low ionization zone is overestimated by the \tii--\tiii\
relation), then the median offsets from the \tii--\tiii\ relation for the
\mstar\ stacks and the \mstar--SFR stacks, respectively, are
\begin{itemize}
\item \teoii: $-$2000~K and $-$1300~K,
\item \tenii: $-$1200~K and $-$1400~K,
\item \tesii: $-$4100~K and $-$3300~K.
\end{itemize}
The \teoii\ and \tenii\ offsets from the \tii--\tiii\ relation for the \mstar\
stacks are consistent given the scatter, which suggests that the \tii--\tiii\
relation overestimates the low ionization zone \te\ by $\sim$1000--2000~K.  The
\tesii\ measurements show a larger offset from the \tii--\tiii\ relation than
\teoii\ and \tenii.  The outlier in the \mstar--SFR \tesii--\teoiii\ panel also
has a high \teoii, but this outlier just corresponds to a single galaxy, so it
may not be representative of all galaxies with this stellar mass and SFR.

The offset between the electron temperatures of the stacks and the \tii--\tiii\
relation is analogous to the trend for individual galaxies found by
\citet{pilyugin2010}, which persists when these galaxies are stacked (see
Section \ref{sec:gal_comp} and Figure \ref{fig:te_p10}a).  The similar
distributions of stacks and individual galaxies relative to the \tii--\tiii\
relation shows that the offset for the stacks is not a by-product of stacking
but rather a reflection of the properties of the individual galaxies (for
further discussion see Section \ref{sec:gal_comp}).

At high SFRs (SFR$_{1.0}^{1.5}$ and SFR$_{1.5}^{2.0}$), the offset in \teoii\
disappears, and the median \teoii\ of these stacks is consistent with the
\tii--\tiii\ relation, albeit with a large dispersion.  The emission from these
galaxies is likely dominated by young stellar populations, whose hard ionizing
spectrum may be similar to the single stellar spectra used by
\citet{stasinska1982} to model \hii\ regions.  However, a single stellar
effective temperature may not be appropriate for galaxy spectra that include a
substantial flux contribution from older \hii\ regions that have softer ionizing
spectra \citep{kennicutt2000, pilyugin2010}.

\begin{figure}
\centerline{\includegraphics[width=9cm]{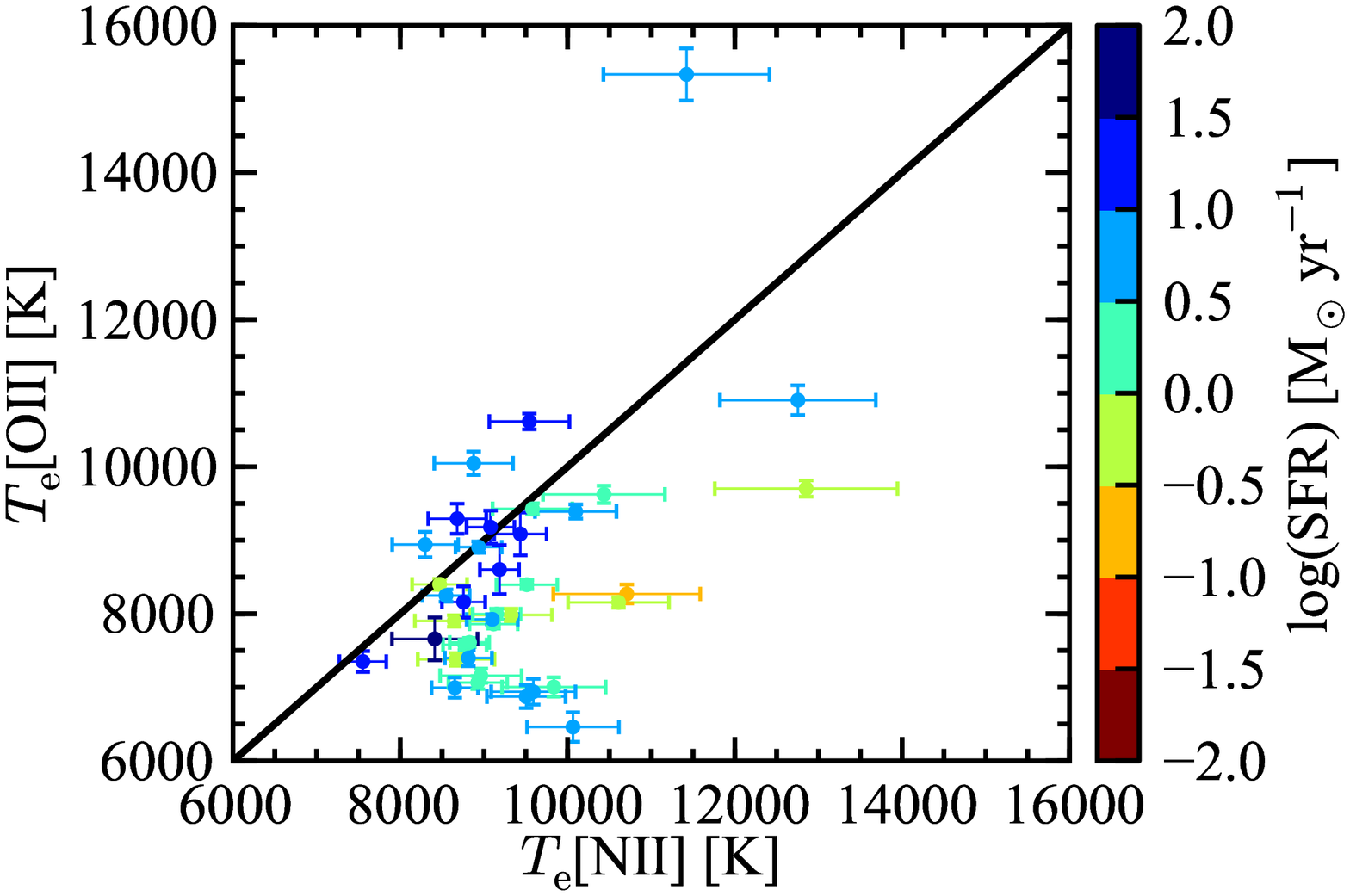}}
\caption{Electron temperatures derived from the \oiis\ line ratios as a function
of electron temperature derived from the \niis\ line ratios for the \mstar--SFR
stacks (color-coded by SFR).  The line indicates \teoii~=~\tenii\ (as assumed in
Equation \ref{eqn:t2t3}).
\label{fig:te_fmr_o2n2}}
\end{figure}

Figure \ref{fig:te_fmr_o2n2} compares the electron temperatures of \oiis\ and
\niis\ for the \mstar--SFR stacks (color-coded by SFR).  In the two-zone model,
both \teoii\ and \tenii\ represent the temperature of the low ionization zone,
so these temperatures should be the same.  The stacks scatter around the line of
equality (black line), though the median offset from the \teoii~=~\tenii\
relation is 1100 K towards higher \tenii.  If only the stacks that also have
detectable \oiiia\ are considered (most of which have \teoii\ $\gtrsim$ 8000 K),
then the median offset from the relation is smaller than the median uncertainty
on \tenii.  The agreement between \teoii\ and \tenii\ for this subset of stacks
is consistent with the similar offsets found for \teoii\ and \tenii\ relative to
the \tii--\tiii\ relation in Figure \ref{fig:te_stacks}.

\subsection{Ionic and Total Abundances}
\label{sec:abund}

We calculated the ionic abundance of O$^+$ and O$^{++}$ with the
\textit{nebular.ionic} routine \citep{derobertis1987, shaw1995} in {\sc
iraf/stsdas}, which determines the ionic abundance from the electron
temperature, electron density, and the flux ratio of the strong emission line(s)
relative to \hbeta.  We derived the ionic abundance uncertainties with the same
Monte Carlo simulations used to compute the electron temperature and density
uncertainties (see Section \ref{sec:te}); the ionic abundance uncertainties were
propagated analytically to calculate the total abundance uncertainties.  We do
not attempt to correct for systematic uncertainties in the absolute abundance
scale.

\begin{figure}
\centerline{\includegraphics[width=9cm]{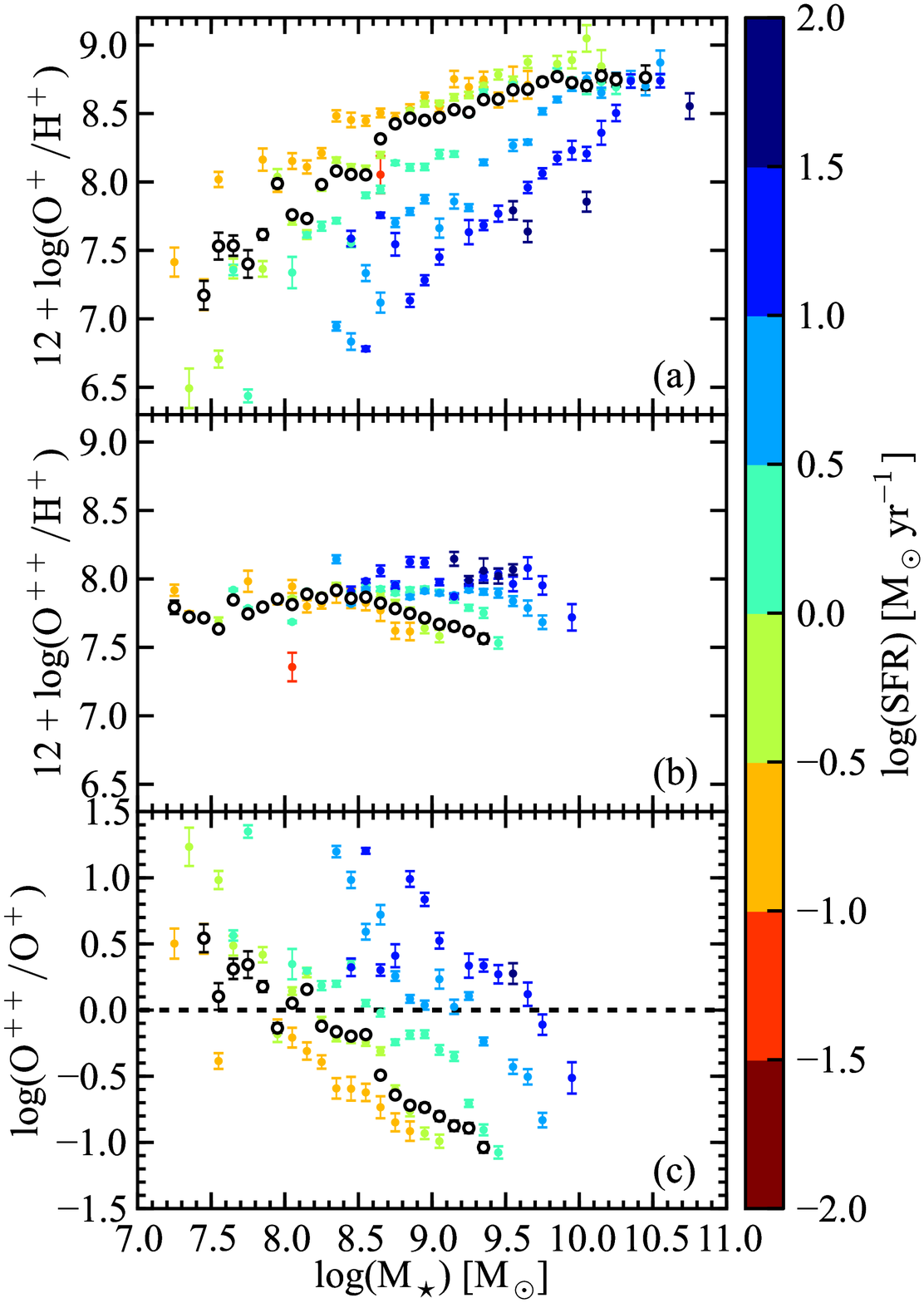}}
\caption{The ionic abundance of O$^+$ (panel a), the ionic abundance of O$^{++}$
  (panel b), and the relative ionic abundance of O$^{++}$ and O$^+$ (panel c) as
  a function of stellar mass for the \mstar\ stacks (open circles) and
  \mstar--SFR stacks (circles color-coded by SFR).  The dashed line in panel (c)
  indicates equal abundances of O$^{++}$ and O$^+$.
\label{fig:o3o2}}
\end{figure}

The top two panels of Figure \ref{fig:o3o2} show the ionic abundance of O$^+$
and O$^{++}$ as a function of stellar mass for the \mstar\ stacks (open circles)
and the \mstar--SFR stacks (circles color-coded by SFR).  The O$^+$ abundance
increases with stellar mass at fixed SFR and decreases with SFR at fixed stellar
mass.  The abundance of O$^{++}$ is relatively constant as a function of stellar
mass but is detected in galaxies with progressively higher SFRs as stellar mass
increases.

In Figure \ref{fig:o3o2}c, we plot the logarithmic ratio of the O$^{++}$ and
O$^+$ abundances as a function of stellar mass.  The dotted line in Figure
\ref{fig:o3o2}c shows equal abundances of O$^+$ and O$^{++}$.  The contribution
of O$^+$ to the total oxygen abundance increases with stellar mass at fixed SFR
and decreases with SFR at fixed stellar mass (i.e., in the same sense as how the
O$^+$ abundance changes with \mstar\ and SFR).  The \textit{O$^+$ abundance
dominates the total oxygen abundance in the majority of the stacks} (i.e., above
log[\mstar]~=~8.2 for the \mstar\ stacks and in half of the \mstar--SFR stacks
with detected \oiia).  Furthermore, the O$^+$ abundance can be measured in many
high stellar mass and/or low SFR stacks that lack a measured O$^{++}$ abundance,
which indicates that O$^+$ is very likely the main ionic species of oxygen in
these stacks too.  A simple extrapolation of the log(O$^{++}$/O$^+$) ratio to
higher stellar masses for the \mstar\ stacks shows that the O$^{++}$ abundance
would contribute less than 10\% of the total oxygen abundance.

\begin{figure}
\centerline{\includegraphics[width=9cm]{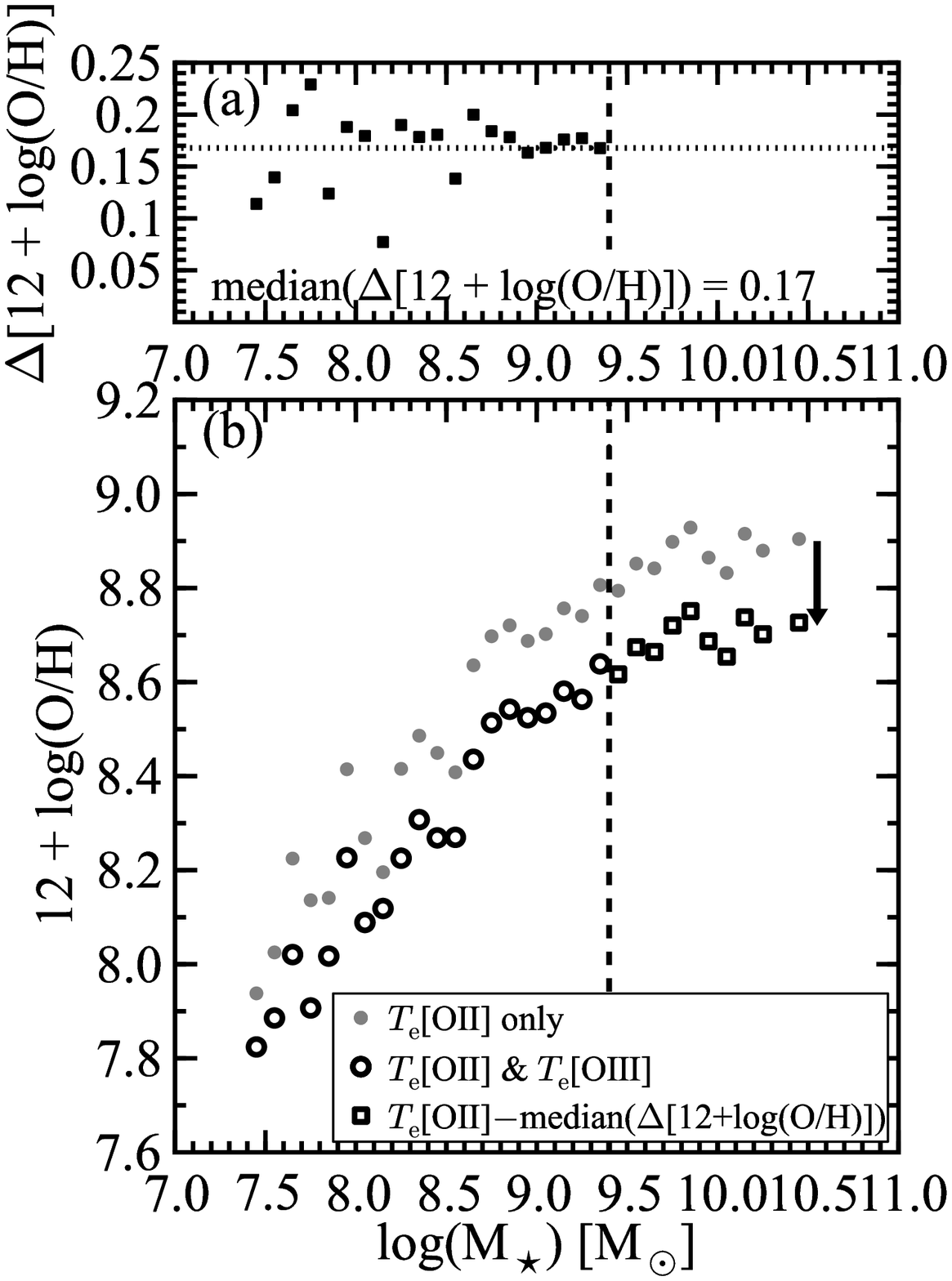}}
\caption{Panel (a): the difference in the direct method metallicity determined
  from \teoii\ only (\teoiii\ was inferred with the \tii--\tiii\ relation given
  in Equation \ref{eqn:t2t3}) and the direct method metallicity determined from
  both \teoii\ and \teoiii.  The dotted line denotes the median difference, and
  the dashed line marks the upper mass cutoff for which \teoiii\ can be
  independently measured in the \mstar\ stacks.  Panel (b): the mass-metallicity
  relation for direct method metallicities determined from \teoii\ only (gray
  circles) and from both \teoii\ and \teoiii\ (open circles).  To account for
  the overestimated metallicity (and underestimated \teoiii) caused by assuming
  the \tii--\tiii\ relation (Equation \ref{eqn:t2t3}), we subtract the median
  metallicity difference shown in panel (a) from the \teoii-based metallicities
  above log(\mstar)~=~9.4 (shown by the dashed line), which results in the open
  squares.  The arrow marks this shift.  The sequence of open circles and
  squares shows the composite direct method metallicities of the \mstar\ stacks
  that we will adopt for the rest of the paper.  We repeated the same procedure
  for each SFR bin of the \mstar--SFR stacks.  The median metallicity
  differences are given in Table \ref{table:mzr_fits}.
\label{fig:mzr_composite}}
\end{figure}

We assume that the total oxygen abundance is the sum of the ionic abundances of
the two dominant species,
\begin{equation}
\frac{\rm O}{\rm H} = \frac{\rm O^+}{\rm H^+} + \frac{\rm O^{++}}{\rm H^+},
\end{equation}
and the total abundance uncertainties were determined by propagating the ionic
abundance uncertainties.  In highly ionized gas, oxygen may be found as
O$^{3+}$, but its contribution to the total oxygen abundance is minimal.
Abundance studies that use the direct method typically measure \teoiii\ and the
O$^{++}$ abundance but adopt the \tii--\tiii\ relation to infer \teoii\ and the
O$^+$ abundance.  However, Figure \ref{fig:te_stacks} shows that the
\tii--\tiii\ relation overestimates \teoii, which leads to an underestimate of
the O$^+$ abundance and the total oxygen abundance.  Many of the stacks have
measured O$^+$ and O$^{++}$ abundances, so the total oxygen abundance can be
measured accurately in these stacks without using the \tii--\tiii\ relation.

To extend our total oxygen abundance measurements to higher stellar mass, we
form a ``composite'' metallicity calibration (see Figure
\ref{fig:mzr_composite}) that uses the O$^+$ and O$^{++}$ abundances when
available and the O$^+$ abundance plus the O$^{++}$ abundance inferred with the
\tii--\tiii\ relation if \teoii\ is measured but not \teoiii\ (in the opposite
sense from how it is normally applied).  The total oxygen abundance of the
latter group of stacks is dominated by the O$^+$ abundance, so the inferred
O$^{++}$ abundance makes only a small contribution (<10\% based on the trend
indicated by Figure \ref{fig:o3o2}c).  A simple combination of these two
metallicity calibrations would lead to a discontinuity at their interface (in
the MZR) because applying the \tii--\tiii\ relation underestimates \teoiii\ and
thus overestimates the O$^{++}$ abundance.  To account for this effect, we
decrease the total oxygen abundances that adopt the \tii--\tiii\ relation by the
median offset between the two calibrations where they are both measured (0.18
dex for the \mstar\ stacks).  For the \mstar--SFR stacks, we calculate the
median offset for each SFR bin (reported in Table \ref{table:mzr_fits}).  The
offsets are nearly constant as a function of \mstar\ and stem from the
approximately constant offset in the \teoii--\teoiii\ plot (top row of Figure
\ref{fig:te_stacks}).  Because we account for the systematic offset from the
\tii--\tiii\ relation, our composite metallicities are insensitive to the exact
choice of the \tii--\tiii\ relation.  The metallicities of the stacks are
presented in Table \ref{table:metallicity}.

Most direct method metallicity studies measure the \oiiia\ line flux but not the
\oiia\ line fluxes, so they must adopt a \teoii--\teoiii\ relation, such as the
\tii--\tiii\ relation, to estimate the O$^+$ abundance.  One reason for this is
the large wavelength separation between the \oii\ strong line and the \oiia\
auroral lines used to measure \teoii.  The flux ratio of these two line
complexes can be affected by a poor reddening correction, particularly for the
\oii\ line, and some spectrographs cannot observe this entire wavelength range
efficiently.  In individual spectra, the \oiia\ lines can be overwhelmed by the
noise, which can lead to a large scatter in the \teoii--\teoiii\ diagram (see
Figure 1 of \citealt{kennicutt2003} or Figure 4 of \citealt{izotov2006}).
Fortunately, the noise near \oiia\ appears to be random and is effectively
reduced by stacking, even without the stellar continuum subtraction (see Figure
\ref{fig:sample_spec}f).  \citet{kennicutt2003} noted that the \oiia\ line
fluxes may be affected by recombination of O$^{++}$, although they find that the
typical contribution to the \oiia\ line fluxes is <5\% (based on the correction
formulae from \citealt{liu2000}) and that \teoii\ is affected by $\sim$2--3\%,
which corresponds to <400~K for the \hii\ regions in their study.

We also calculated the ionic abundance of N$^+$ with {\it nebular.ionic},
similar to the procedure used to calculate the ionic abundances of O$^+$ and
O$^{++}$, except that we adopt \teoii\ as the electron temperature instead of
\tenii\ because the \oiia\ lines are detected in more stacks and with higher SNR
than the \niia\ line (see Figure \ref{fig:sample_spec}).  The relative ionic
abundance of N$^+$/O$^+$ was derived from the ionic abundances of each species.
We then assume that N/O~=~N$^+$/O$^+$ \citep{peimbert1969, garnett1990} to
facilitate comparison with other studies in the literature
\citep[e.g.,][]{vilacostas1993}.  Although this assumption is uncertain,
\citet{nava2006} found that it should be accurate to $\sim$10\% for low
metallicity objects (12~+~log[O/H]~$\leq$~8.1).  The N/O abundances of the
stacks are reported in Table \ref{table:metallicity}.

\subsection{Strong Line Metallicities}
\label{sec:sl_abund}
We compare our direct method metallicities to strong line metallicities with
various empirical and theoretical calibrations of the most common line ratios:
\begin{itemize}
\item \rtt: (\oii~+~\oiii)~/~\hbeta
\item N2O2: \niir~/~\oii, 
\item N2: \niir~/~\halpha, 
\item O3N2: (\oiiir~/~\hbeta)~/~(\niir~/~\halpha).
\end{itemize}
We derived metallicities for our stacks with the theoretical \rtt\ calibrations
of \citet[hereafter M91]{mcgaugh1991}, \citet[hereafter Z94]{zaritsky1994}, and
\citet[hereafter KK04]{kobulnicky2004}; the hybrid empirical--theoretical N2
calibration of \citet[hereafter D02]{denicolo2002}; the theoretical N2O2
calibration of \citet[hereafter KD02]{kewley2002}; and the mostly empirical N2
and O3N2 calibrations of \citet[hereafter PP04]{pettini2004}.  We determined
uncertainties on the strong line metallicities with the Monte Carlo simulations
detailed in Section \ref{sec:te}; these uncertainties do not account for
systematic uncertainties in the absolute abundance scale.  For a detailed
discussion of these calibrations and formulae to convert between the
metallicities derived from each calibration see \citet{kewley2008}.


\section{How Does Stacking Affect Measured Electron Temperatures and Metallicities?}
\label{sec:gal_comp}

\begin{figure}
\centerline{\includegraphics[width=9cm]{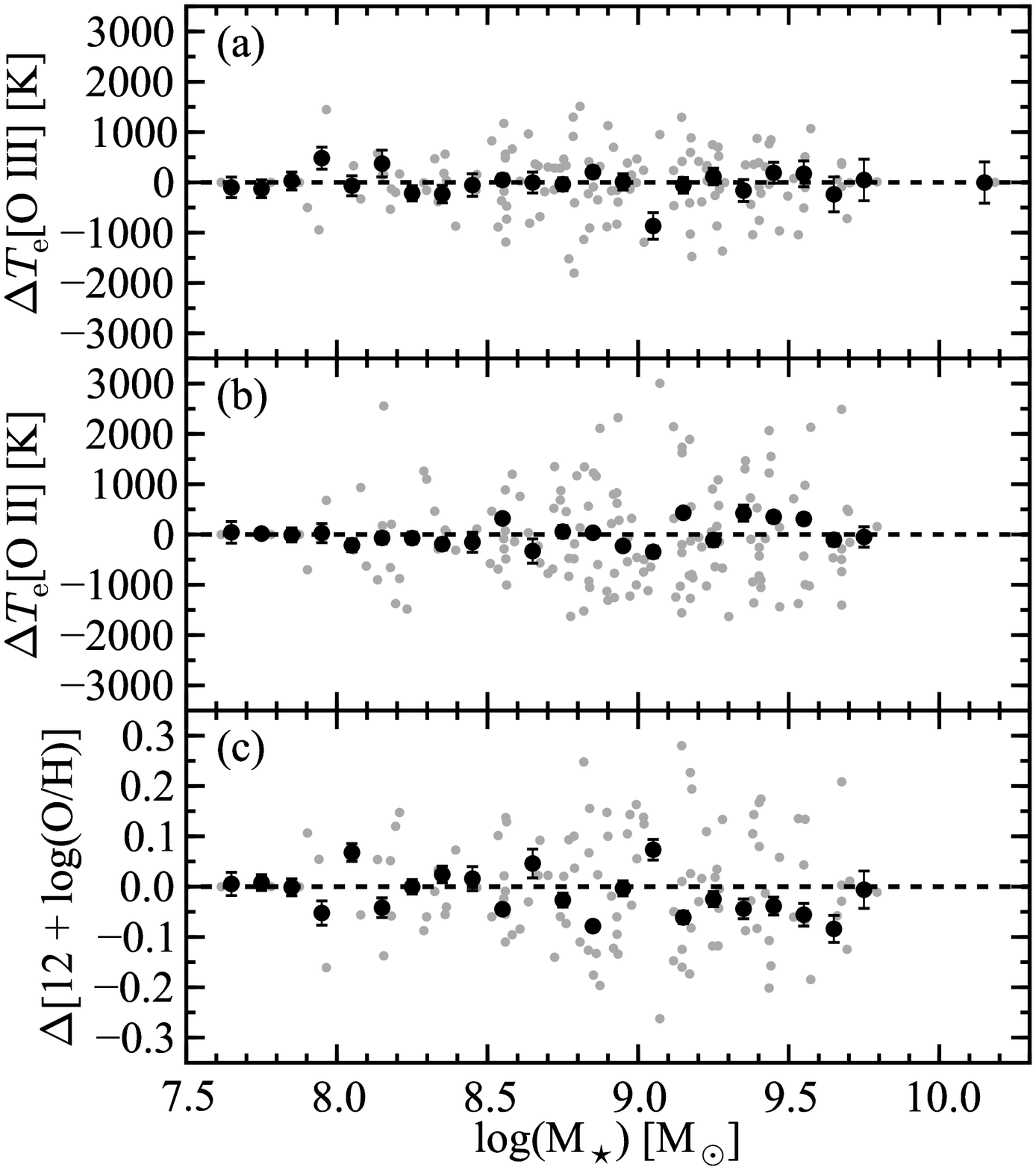}}
\caption{\teoiii, \teoii, and direct method metallicity for individual spectra
  (small gray circles) and stacks in bins of 0.1 dex in stellar mass (large
  black circles) for the \citet{pilyugin2010} sample relative to the mean of
  galaxies within a stellar mass bin of width 0.1 dex \msun\ (shown by the
  dashed line in each panel).  The stacks are consistent with the mean \teoiii,
  \teoii, and metallicity within the measurement uncertainties.
\label{fig:pilyugin_comp}}
\end{figure}

\begin{figure}
\centerline{\includegraphics[width=9cm]{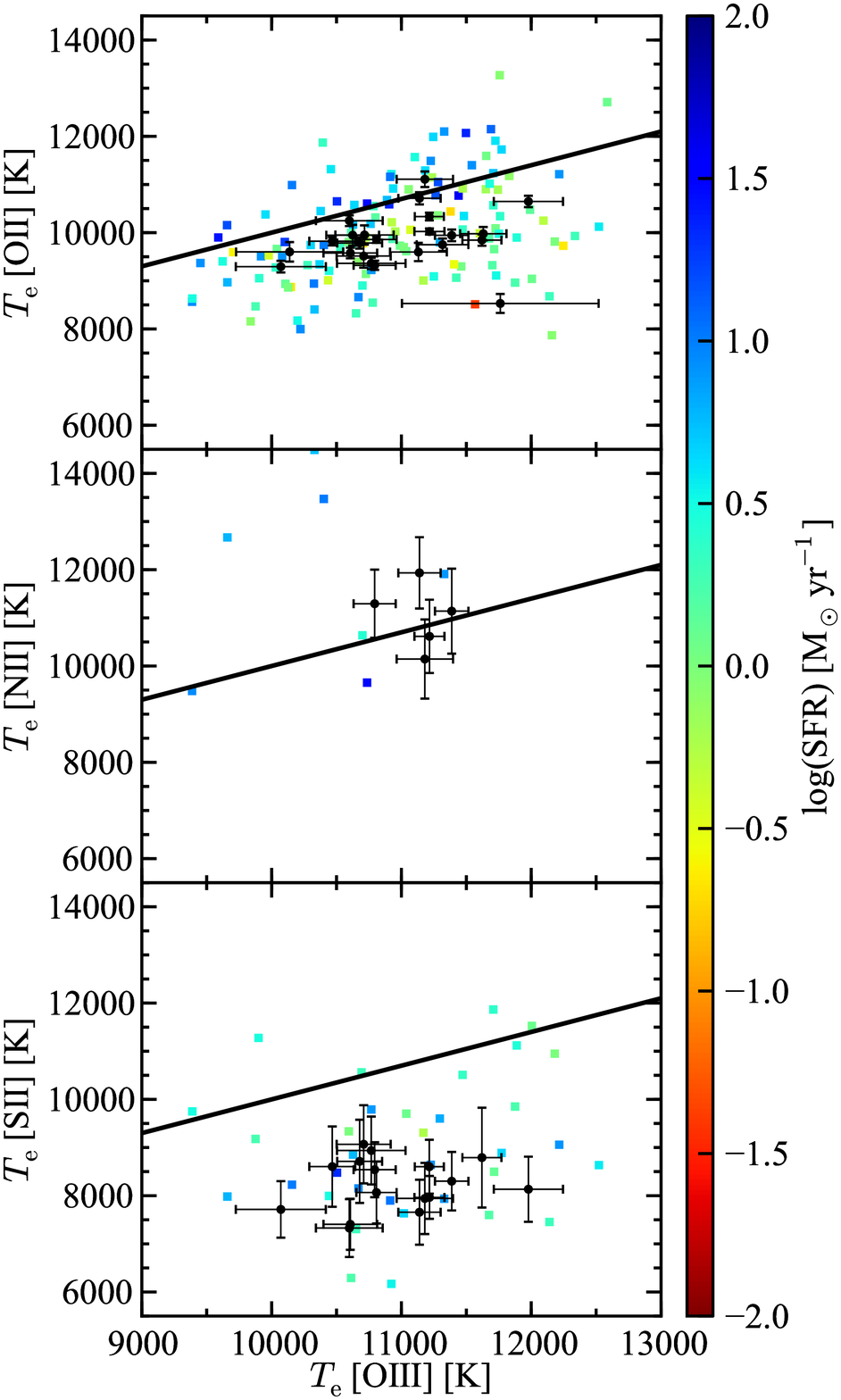}}
\caption{The electron temperatures derived from the \oiis, \niis, and \siis\
  line ratios as a function of electron temperature derived from the \oiiis\
  line ratio for the \citet{pilyugin2010} sample of galaxies with detectable
  \oiiia\ and \oiia\ (squares color-coded by SFR; see Section
  \ref{sec:gal_comp}) and stacks of the same galaxies in bins of 0.1 dex in
  stellar mass (black circles).  The black line shows the \tii--\tiii\ relation
  (Equation \ref{eqn:t2t3}).
\label{fig:te_p10}}
\end{figure}

Stacking greatly increases SNR and thus enables measurements of physical
properties that are unattainable for individual objects.  However, measurements
from stacked spectra are only meaningful if they represent the typical
properties of the objects that went into the stack.  To evaluate the effect of
stacking on the electron temperatures and metallicities, we stacked a sample of
181 SDSS DR6 \citep{adelmanmccarthy2008} galaxies with individual detections of
\oiiia\ and \oiia\ from \citet{pilyugin2010} in bins of 0.1 dex in stellar mass.
Figure \ref{fig:pilyugin_comp} shows the \teoiii, \teoii, and the direct method
metallicities of the individual galaxies (gray squares) and stacks (black
circles) relative to the mean of the galaxies that went into each stack.  For
all three properties, the stacks are consistent with the mean of the galaxies
within the measurement uncertainties, which demonstrates that the properties
derived from galaxies stacked in narrow bins of stellar mass are representative
of the mean properties of the input galaxies.

In Figure \ref{fig:te_p10}, the \oiis, \niis, and \siis\ electron temperatures
are plotted as a function of the \oiiis\ electron temperature for the galaxies
(squares color-coded by SFR) and stacks (black circles).  The black line in each
panel indicates the \tii--\tiii\ relation (Equation \ref{eqn:t2t3}).  The stacks
fall within the distribution of galaxies in the \teoii--\teoiii\ and
\tesii--\teoiii\ plots (Figure \ref{fig:te_p10}a,c).  There is some discrepancy
between the stacks and galaxies in the \tenii--\teoiii\ plot (Figure
\ref{fig:te_p10}b), but the paucity of \niia\ detections limits the usefulness
of any strong conclusions based on \tenii.  Overall, the qualitative agreement
between the electron temperatures of the stacks and galaxies, especially for
\teoii\ and \tesii, demonstrates that the offset from the \tii--\tiii\ relation
for the stacks shown in Figure \ref{fig:te_stacks} is not an artifact of
stacking.

The majority of the galaxies lie below the \tii--\tiii\ relation, as was
previously shown \citet{pilyugin2010}.  We find a similar result for the
galaxies in the \tesii--\teoiii\ relation.  Galaxies with moderate SFRs
(log[SFR]~$\sim$~0.0) are preferentially further below the \tii--\tiii\ relation
than galaxies with high SFRs (log[SFR]~$\gtrsim$~1.0) in the \teoii--\teoiii\
plot.  A similar effect is also present in the \mstar--SFR stacks.
\citet{pilyugin2010} found that galaxies with lower excitation parameters and
\oiiir/\hbeta\ flux ratios had larger offsets from the \tii--\tiii\ relation,
which is consistent with our result based on SFR.  They showed that the offset
from the \tii--\tiii\ relation is likely due to the combined emission from
multiple ionizing sources by comparing the observed \teoii--\teoiii\ relation
with the temperature predicted by \hii\ region models that include ionizing
sources of various temperatures.  Based on these models, they concluded that
differences in the hardness of the ionizing radiation, caused by the
age-dependence of \hii\ region spectral energy distributions, govern the scatter
in the \teoii--\teoiii\ plot for their sample of galaxies.  Both our results and
theirs suggest that galaxies with higher SFRs are more similar to the \hii\
region models that served as the basis for the \tii--\tiii\ relation than
galaxies with moderate SFRs.  This is because they are more likely to be
dominated by younger stellar populations that are better approximated by the
input to the \citet{stasinska1982} models (see Section \ref{sec:te} for
additional discussion).

The electron temperatures and metallicities of the stacks are unbiased relative
to those of the input galaxies, but there is some evidence that the integrated
galaxy electron temperature and metallicity are systematically higher and
lower, respectively, than the electron temperatures and metallicities of the
individual \hii\ regions in the galaxy.  \citet{kobulnicky1999} compared the
electron temperatures and metallicities of individual \hii\ regions in a galaxy
to the pseudo-global values derived by stacking the spectra of the individual
\hii\ regions.  They showed that the electron temperatures and direct method
metallicities of their galaxies were biased towards higher temperatures and
lower metallicities by $\sim$1000--3000~K and 0.05--0.2 dex, respectively,
relative to the median values of the individual \hii\ regions.  Global spectra
are biased because they are the luminosity-weighted average of the \hii\
regions, whose properties can vary widely (see, e.g., the large scatter around
the \tii--\tiii\ relation for \te\ measurements of individual \hii\ regions in
Figure 1 of \citealt{kennicutt2003} or Figure 4 of \citealt{izotov2006}).  The
fluxes of the auroral lines might be particularly affected by a
luminosity-weighted average because auroral line flux decreases non-linearly
with metallicity.  While \citet{kobulnicky1999} only studied the effects on
\oiiia, the relative contribution of each \hii\ region likely varies among the
commonly measured ionic species, potentially yielding results that do not agree
with the \tii--\tiii\ relation.  We also note that their method of stacking
\hii\ regions does not perfectly simulate global line flux measurements because
it does not account for the contribution of diffuse ionized gas (i.e., the
emission from gas not in \hii\ regions), which may affect the \niis\ and \siis\
line fluxes \citep[see][]{moustakas2006}.  In summary, the differences in
electron temperatures and metallicities between galaxy stacks and individual
\hii\ regions are dominated by the systematic offset between global galaxy
properties and individual \hii\ regions rather than any effects from stacking
the global galaxy spectra.

\begin{figure}
\centerline{\includegraphics[width=9cm]{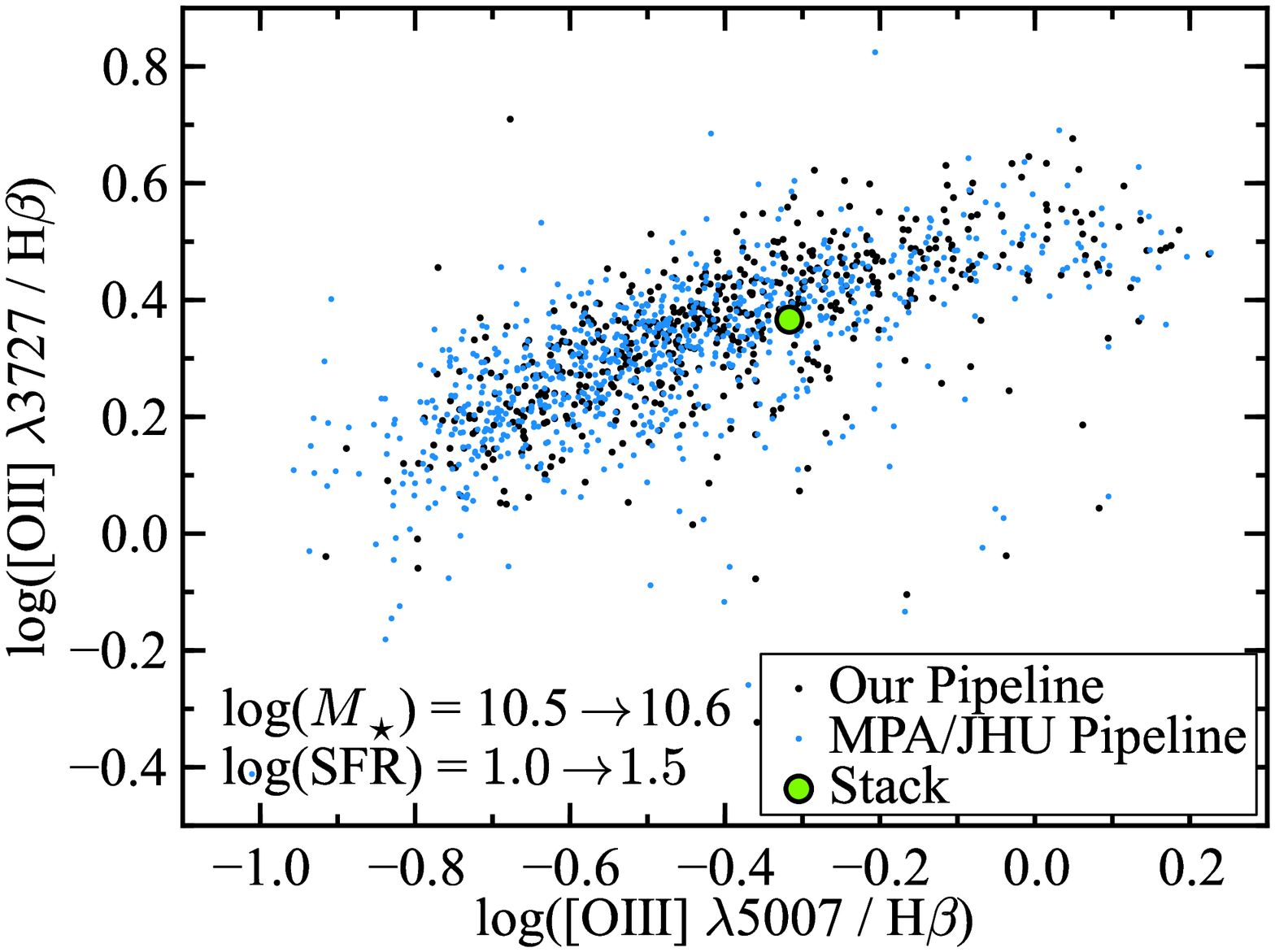}}
\caption{\oii\ and \oiiir\ fluxes relative to \hbeta\ of galaxies in one
  \mstar--SFR bin (log[\mstar]~=~10.5--10.6 and log[SFR]~=~1.0--1.5) and the
  stack of those galaxies.  The small black and blue circles represent
  individual galaxies with fluxes measured with our pipeline and the MPA-JHU
  pipeline \citepalias{tremonti2004}, respectively.  The large green circle
  corresponds to the stack of the same galaxies.
\label{fig:O2O3}}
\end{figure}

The auroral lines are undetectable in high stellar mass galaxies, so we
investigate the effect of stacking by comparing the oxygen strong line fluxes of
individual galaxies to the stack of those galaxies.  Figure \ref{fig:O2O3} shows
the \oii\ and \oiiir\ fluxes relative to \hbeta\ for individual galaxies (small
black and blue circles) with log(\mstar)~=~10.5--10.6 and log(SFR)~=~1.0--1.5
and the stack of the same galaxies (large green circle).  The small black and
blue circles correspond to the line fluxes determined with our pipeline and the
MPA-JHU pipeline, respectively.  The distribution of individual galaxies with
fluxes measured by our pipeline and the MPA-JHU pipeline coincide well.  In
detail, the median fluxes from our pipeline are 0.08 and 0.04 dex higher for
\oiiir\ and \oii, respectively, than the median fluxes from the MPA-JHU
pipeline.  The \oiiir\ and \oii\ fluxes of the stack are 0.09 and 0.01 dex
higher, respectively, than the median of fluxes from our pipeline.  Although the
spread is large in the individual galaxies (>1 dex for both \oiiir\ and \oii),
the stack is representative of the typical line fluxes of individual galaxies
that went into the stack.

We also note that many of our stacks contain far more galaxies than are needed
to simply detect a given line, and thus are unlikely to be dominated by a few,
anomalous galaxies.  As an example, we estimate how many galaxies would need to
be stacked for a detection of \oiia.  If we assume that the uncertainty on the
line flux decreases as $\sqrt{N_\mathrm{galaxies}}$, the error on the
measurement of any individual galaxy is $\sigma_\mathrm{stack} *
\sqrt{N_\mathrm{galaxies}}$.  We use a 5$\sigma$ detection threshold, so the
minimum number of galaxies needed to detect a line is $N = [(5\sigma) /
\mathrm{flux}]^2$.  For the \mstar$_{9.5}^{9.6}$--SFR$_{0.0}^{0.5}$ stack, the
minimum number of galaxies required to detect \oiia\ is
$N_\mathrm{galaxies}$~=~40, which is well below the actual number of galaxies
(1996) in this stack.


\section{The Mass--Metallicity Relation and Mass--Metallicity--SFR Relation}
\label{sec:mzr_fmr}

\subsection{The Mass--Metallicity Relation}
\label{sec:mzr}

\begin{figure*}
\centerline{\includegraphics[width=19cm]{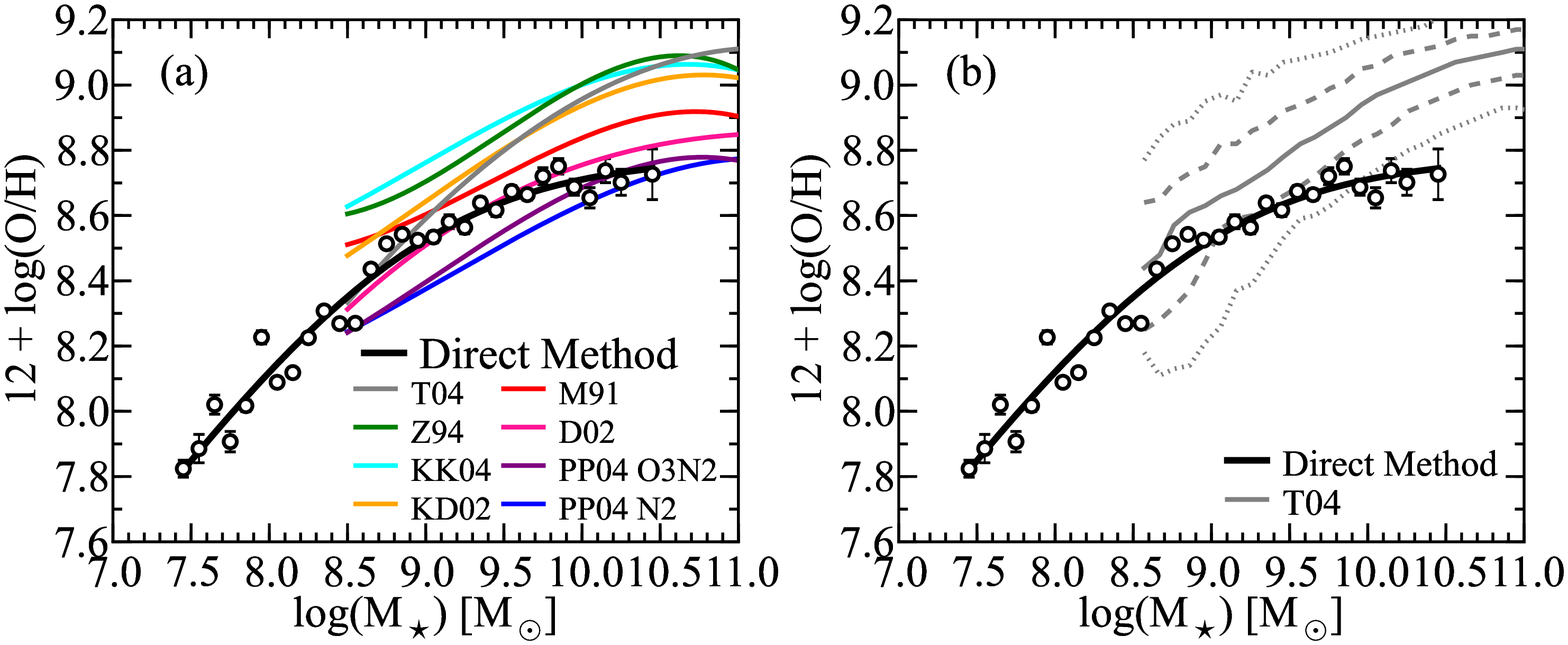}}
\caption{The direct method mass--metallicity relation for the \mstar\ stacks
  (circles).  In both panels, the thick black solid line shows the asymptotic
  logarithmic fit to the direct method measurements (see Equation
  \ref{eqn:alog}).  Panel (a): the colored lines represent various strong line
  calibrations (\citealt{tremonti2004}; \citealt{zaritsky1994};
  \citealt{kobulnicky2004}; \citealt{kewley2002}; \citealt{mcgaugh1991};
  \citealt{denicolo2002}; \citealt{pettini2004}).  Panel (b): the solid, dashed,
  and dotted gray lines show the median, 68\% contour, and 95\% contour,
  respectively, of the \citet{tremonti2004} MZR.  The metallicities and fit
  parameters for the stacks are reported in Tables \ref{table:metallicity} and
  \ref{table:mzr_fits}, respectively.
\label{fig:mzr}}
\end{figure*}

In Figure \ref{fig:mzr}, we plot the MZR with direct method metallicities for
the \mstar\ stacks (circles).  We fit the MZR for the \mstar\ stacks (black
line) with the asymptotic logarithmic formula suggested by
\citet{moustakas2011}:
\begin{equation}
12 + {\rm log(O/H)} = 12 + {\rm log(O/H)}_{\rm asm} - {\rm log}\left(1 +
\left(\frac{M_{\rm TO}}{M_\star}\right)^\gamma\right),
\label{eqn:alog}
\end{equation}
where 12+log(O/H)$_{\rm asm}$ is the asymptotic metallicity, $M_{\rm TO}$ is the
turnover mass, and $\gamma$ controls the slope of the MZR.  This functional form
is preferable to a polynomial because polynomial fits can produce unphysical
anticorrelations between mass and metallicity, particularly when extrapolated
beyond the mass range over which they were calibrated.  The metallicities and
fit parameters for the stacks are reported in Tables \ref{table:metallicity} and
\ref{table:mzr_fits}, respectively.  For comparison, we show the robust cubic
polynomial fits of eight strong line MZRs (colored lines) from
\citet{kewley2008} in Figure~\ref{fig:mzr}a.  The \citetalias{tremonti2004},
\citetalias{zaritsky1994} \rtt, \citetalias{kobulnicky2004} \rtt,
\citetalias{kewley2002} N2O2, and \citetalias{mcgaugh1991} \rtt\ MZRs are based
on theoretical calibrations, whereas the \citetalias{denicolo2002} N2,
\citetalias{pettini2004} O3N2, and \citetalias{pettini2004} N2 MZRs are based on
empirical calibrations.  In Figure~\ref{fig:mzr}b, the solid, dashed, and dotted
gray lines indicate the median, 68\% contour, and 95\% contour, respectively, of
the \citetalias{tremonti2004} MZR.

\begin{deluxetable*}{cccccc}
\tabletypesize{\small}
\tablecaption{Mass--Metallicity Relation Fit Parameters
\label{table:mzr_fits}}
\tablewidth{0pt}
\tablehead{
\colhead{Stacks} &
\colhead{log($M_\mathrm{TO}$)} &
\colhead{12+log(O/H)$_\mathrm{asm}$} &
\colhead{$\gamma$} &
\colhead{Fit Range} &
\colhead{Median log(O/H) Offset} \\
\colhead{SFR in $M_\odot$ yr$^{-1}$} &
\colhead{[M$_\odot$]} &
\colhead{[dex]} &
\colhead{} &
\colhead{log($M_\star$) [M$_\odot$]} &
\colhead{[dex]} \\
\colhead{(1)} &
\colhead{(2)} &
\colhead{(3)} &
\colhead{(4)} &
\colhead{(5)} &
\colhead{(6)} \\
}
\startdata
MZR  &   8.901  &  8.798   &   0.640  &  7.4--10.5  &  0.18 \\
$-1.0$ $\leq$ log(SFR) < $-0.5$ &   8.253  &  8.726   &   0.734  &  7.2--9.7  &
0.15 \\
$-0.5$ $\leq$ log(SFR) < 0.0  &   9.608  &  9.118   &   0.610  &  7.3--10.2  &  0.13 \\
0.0 $\leq$ log(SFR) < 0.5   &   9.836  &  8.997   &   0.534  &  7.6--10.3  &
0.12 \\
0.5 $\leq$ log(SFR) < 1.0   &   27.225 &  16.383  &   0.449  &  8.3--10.6  &
0.10 \\
1.0 $\leq$ log(SFR) < 1.5   &   32.650 &  16.988  &   0.373  &  8.4--10.6  &
$-$0.04 \\
1.5 $\leq$ log(SFR) < 2.0   &   28.369 &  16.259  &   0.438  &  9.5-10.8  &
0.04 \\
\enddata

\tablecomments{Mass--Metallicity Relation given by
12~+~log(O/H)~=~12~+~log(O/H)$_\mathrm{asm}$~$-$~log(1~+~($M_\mathrm{TO}$~/~$M_\star$)$^\gamma$).
Column (1): Stacks included in fits.  MZR refers to \mstar\ stacks.  The SFR
ranges refer to \mstar--SFR stacks.  Column (2): Turnover mass.  Column (3):
Asymptotic metallicity.  Column (4): Power-law slope.  Column (5): Stellar mass
range of each fit.  Column (6): Median offset between the metallicity determined
with (i) measured \teoii\ and inferred \teoiii\ from the \tii--\tiii\ relation
and (ii) measured \teoii\ and measured \teoiii\ (see Section \ref{sec:abund}).}

\end{deluxetable*}

The most prominent aspect of the direct method MZR is its extensive dynamic
range in both stellar mass and metallicity.  It spans three decades in stellar
mass and nearly one decade in metallicity; this wide range is critical for
resolving the turnover in metallicity with a single diagnostic that is a
monotonic relation between line strength and metallicity.  The broad range in
galaxy properties includes the turnover in the MZR, which is the first time this
feature has been measured with metallicities derived from the direct method.
Our stacked spectra also extend the direct method MZR to sufficiently high
masses that there is substantial overlap with strong line measurements, and we
use this overlap to compare them.

The direct method MZR shares some characteristics with strong line MZRs but
differs in important ways, as can be seen in Figure \ref{fig:mzr}a.  The low
mass end of the direct method MZR starts at log(\mstar)~=~7.4, a full decade
lower than the strong line MZRs.  Nonetheless, naive extrapolations of the
\citetalias{tremonti2004}, \citetalias{denicolo2002}, \citetalias{pettini2004},
and \citetalias{pettini2004} MZRs are in reasonable agreement with our direct
method MZR.  At a stellar mass of log(\mstar)~=~8.5, the lowest stellar mass
where strong line MZRs are reported, the direct method MZR is consistent with
the \citetalias{tremonti2004} and the \citetalias{denicolo2002} MZRs.  Above
this mass, the direct method MZR and the \citetalias{denicolo2002} MZR diverge
from the \citetalias{tremonti2004} MZR.  At log(\mstar)~=~8.9, the direct
method MZR turns over.  By contrast, the strong line MZRs turns over at a much
higher stellar mass (log[\mstar]~$\sim$~10.5): a significant difference that
has implications for how the MZR is understood in a physical context, which we
discuss in Section \ref{sec:physics}.  At high mass, the direct method MZR is
in good agreement with the empirical strong line calibration MZRs, but the
theoretical \citetalias{tremonti2004}, \citetalias{zaritsky1994},
\citetalias{kobulnicky2004}, and \citetalias{kewley2002} strong line
calibration MZRs are offset to higher metallicities by $\sim$0.3 dex at
log(\mstar)~=~10.5, the highest mass stack with detected auroral lines.  Figure
\ref{fig:mzr}b shows the direct method MZR in relation to the scatter of the
\citetalias{tremonti2004} MZR.  The direct method MZR is slightly below the
median \citetalias{tremonti2004} MZR at log(\mstar)~=~8.5, crosses the
16$^\mathrm{th}$ percentile at log(\mstar)~=~9.0, and drops below the
2$^\mathrm{nd}$ percentile at log(\mstar)~=~9.9.  Formally, the direct method
MZR has a scatter of $\sigma$~=~0.05~dex, but this value is not directly
comparable to the scatter in the \citetalias{tremonti2004} MZR because stacking
effectively averages over all galaxies in a bin, which erases information about
galaxy-to-galaxy scatter.

At low masses (log[\mstar]~=~7.4--8.9; i.e., below the turnover), the direct
method MZR scales as approximately O/H~$\propto$~\mstar$^{1/2}$.  While a
comparison over the same mass range is not possible for the
\citetalias{tremonti2004} MZR, its low mass slope, as determined from
log(\mstar)~=~8.5--10.5, is shallower with O/H~$\propto$~\mstar$^{1/3}$.  The
discrepancy in the low mass slopes between the direct method and the
\citetalias{tremonti2004} MZRs could be reasonably explained by the difference
in the mass ranges over which the slopes were measured if the MZR steepens with
decreasing stellar mass (cf., \citealt{lee2006}).  We note that the direct
method and \citetalias{denicolo2002} MZRs have similar slopes and
normalizations over a wide range in masses from log(\mstar)~=~8.5--10.0.

\subsection{Mass--Metallicity--SFR Relation}
\label{sec:mzsfr}

\begin{figure}
\centerline{\includegraphics[width=9cm]{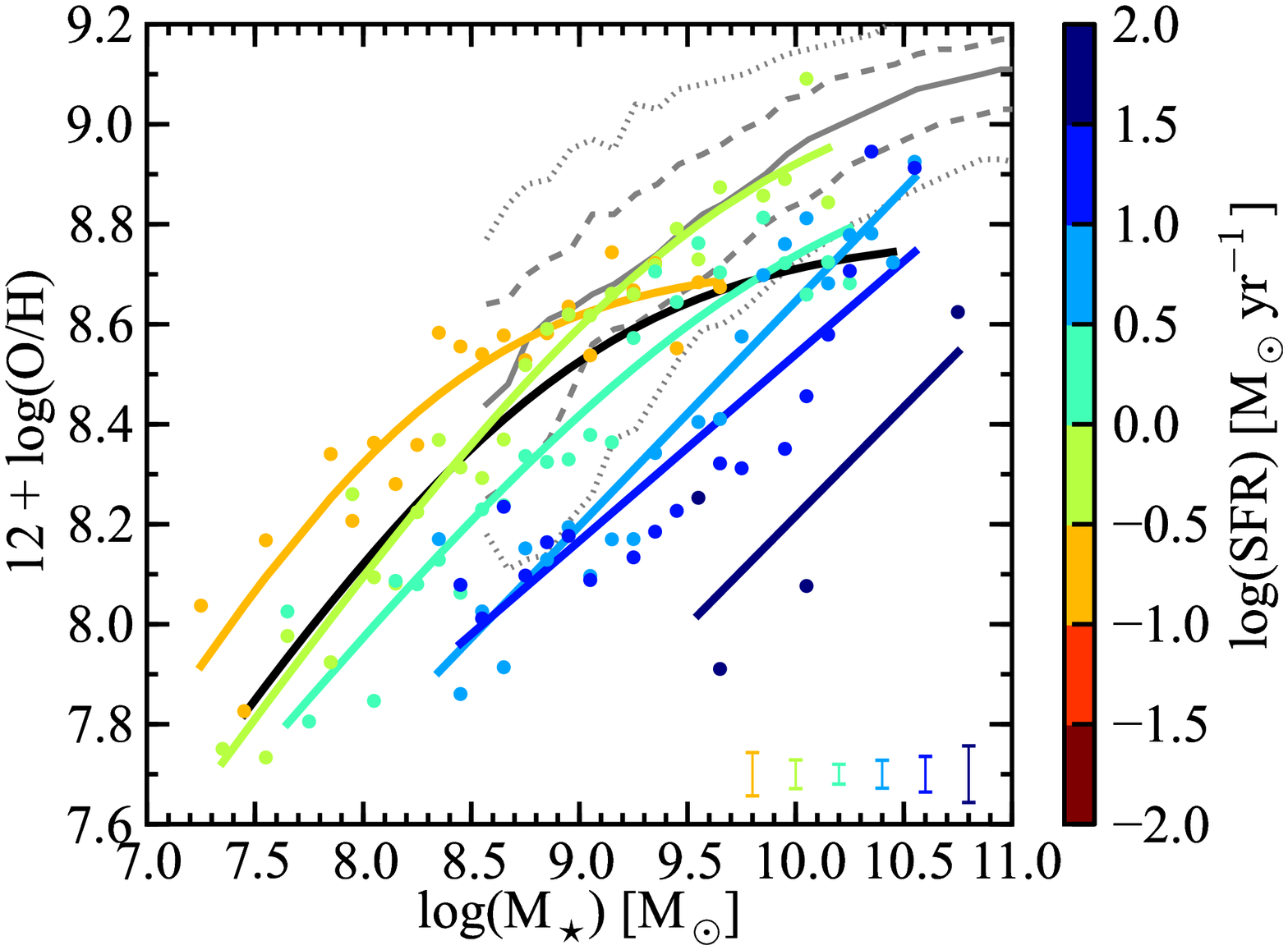}}
\caption{The direct method \mstar--$Z$--SFR relation for the \mstar--SFR stacks
  (circles color-coded by SFR) in the mass--metallicity plane.  The thick solid
  lines color-coded by SFR show the asymptotic logarithmic fits (see Equation
  \ref{eqn:alog}) for the \mstar--SFR stacks.  The thick black line shows the
  direct method MZR from Figure \ref{fig:mzr}.  The solid, dashed, and dotted
  gray lines show the median, 68\% contour, and 95\% contour, respectively, of
  the \citet{tremonti2004} MZR.  The error bars correspond to the mean error for
  the \mstar--SFR stacks of a given SFR.  The metallicities and fit parameters
  for the stacks are given in Tables \ref{table:metallicity} and
  \ref{table:mzr_fits}, respectively.
\label{fig:mzr_sfr}}
\end{figure}

The features of the direct method MZR are shaped by the SFR-dependence of the
MZR, which we investigate with the \mstar--SFR stacks.  Figure \ref{fig:mzr_sfr}
shows the \mstar--SFR stacks (circles color-coded by SFR) in the
mass--metallicity plane (see Figure \ref{fig:msfr} for the number of galaxies
per stack).  The solid colored lines indicate the asymptotic logarithmic fits
(Equation \ref{eqn:alog}) of the \mstar--SFR stacks of a given SFR, hereafter
referred to as SFR tracks (e.g., the orange line is the SFR$_{-1.0}^{-0.5}$
track).  The solid black line is the direct method MZR of the \mstar\ stacks
from Figure \ref{fig:mzr}; the solid, dashed, and dotted gray lines are the
median, 68\% contour, and 95\% contour, respectively, of the
\citetalias{tremonti2004} MZR.  The error bars represent the mean error for the
\mstar--SFR stacks of a given SFR.

The \mstar--SFR stacks help establish the robustness of the direct method MZR.
The low turnover mass and metallicity of the direct method MZR relative to the
\citetalias{tremonti2004} and other theoretical strong line calibration MZRs is
reminiscent of empirical strong line calibration MZRs that suffer from a lack of
sensitivity at high metallicities.  However, the most metal-rich \mstar--SFR
stacks have some of the highest direct method metallicities
(12~+~log[O/H]~>~9.0)---metallicities well above the turnover metallicity of the
direct method MZR.  These measurements unambiguously demonstrate that the
turnover in the direct method MZR is not caused by a lack of sensitivity to high
metallicities.

The \mstar--SFR stacks also can be used to test if galaxies with the highest
SFRs at a given stellar mass disproportionately influence the line fluxes and
metallicities of the \mstar\ stacks.  High SFR galaxies have more luminous
emission lines and lower metallicities and thus may dominate the inferred
metallicity of the stack.  To investigate this possibility, we calculated the
difference between the metallicity of the \mstar\ stack and the galaxy
number-weighted average of the metallicities of the \mstar--SFR stacks (for the
stacks with measured metallicities) at a given stellar mass.  The median offset
is only $-$0.037 dex in metallicity; for reference, the median metallicity
uncertainties for the \mstar\ and \mstar--SFR stacks are 0.019 and 0.027 dex,
respectively.  The slight offset could be due to preferentially including the
metallicities of \mstar--SFR stacks with higher SFR (lower metallicity) relative
to lower SFR (higher metallicity) in the weighted average because the former
tend to have larger line fluxes than the latter, whereas the \mstar\ stacks
include the contribution from galaxies of all SFRs at a given stellar mass.
Still, the magnitude of this offset is small, which indicates that the highest
SFR galaxies do not have an appreciable effect on the metallicity of the \mstar\
stacks because they are quite rare (see Figure \ref{fig:msfr}).  Furthermore,
the metallicities of the \mstar\ stacks effectively track the metallicity of the
most common galaxies at a given stellar mass.

The most striking features of Figure \ref{fig:mzr_sfr} are the 0.3--0.6 dex
offsets in metallicity at fixed stellar mass between the \mstar--SFR stacks.
This trend results from the substantial, nearly monotonic dependence of the MZR
on SFR.  At a given stellar mass, higher SFR stacks almost always have lower
metallicities than lower SFR stacks, so there is little overlap between the
different SFR tracks.  Furthermore, the small regions with overlap may be the
result of the observational uncertainties.

The interplay between stellar mass, SFR, and metallicity for typical galaxies is
reflected in the features of the direct method MZR, especially the turnover
mass.  The constant SFR tracks (colored lines in Figure \ref{fig:mzr_sfr}) show
that metallicity increases with stellar mass at fixed SFR.  However, the typical
SFR also increases with stellar mass, which shifts the ``typical'' galaxy (as
measured by the \mstar\ stacks) to progressively higher SFR and consequently
lower metallicity at fixed stellar mass.  Taken together, the turnover in the
MZR is the result of the conflict between the trend for more massive galaxies to
have higher SFRs and the trend for metallicity to decrease with SFR at fixed
mass.  The turnover in the \citetalias{tremonti2004} MZR (and other strong line
calibration MZRs) occurs at a higher stellar mass than the the direct method MZR
because the strong line metallicity calibrations produce a weaker
SFR--metallicity anticorrelation.  This means that the progression to higher
SFRs with increasing stellar mass has less of an effect on the MZR.

Interestingly, the SFR$_{-0.5}^{0.0}$ stacks (light green circles/line) are
nearly identical to the \citetalias{tremonti2004} MZR in slope, shape, turnover,
and normalization.  While the exact cause of this agreement is unclear, it is
possible that the photoionization models that underlie the
\citetalias{tremonti2004} metallicities assume physical parameters that are most
appropriate for galaxies with this range of SFR.  We discuss potential
systematic effects of strong line calibrations in Section \ref{sec:alpha}.

The stacks with very high SFRs (SFR$_{1.0}^{1.5}$ and SFR$_{1.5}^{2.0}$; blue
and dark blue circles/lines, respectively) have significantly lower
metallicities than the stack of all galaxies at fixed mass in the MZR.  The high
SFRs and low metallicities of these galaxies suggests that they are probably
undergoing major mergers, as found by \citet{peeples2009} for similar outliers.
Major mergers drive in considerable amounts of low metallicity gas from large
radii, which dilutes the metallicity of the galaxy and triggers vigorous star
formation \citep[e.g.,][]{kewley2006, kewley2010, torrey2012}.  These stacks
also have a larger scatter than lower SFR stacks, which is likely driven by the
small numbers of galaxies per stack coupled with the large intrinsic dispersion
in the individual galaxy metallicities.

\subsection{The Fundamental Metallicity Relation}
\label{sec:fmr}

\begin{figure}
\centerline{\includegraphics[width=9cm]{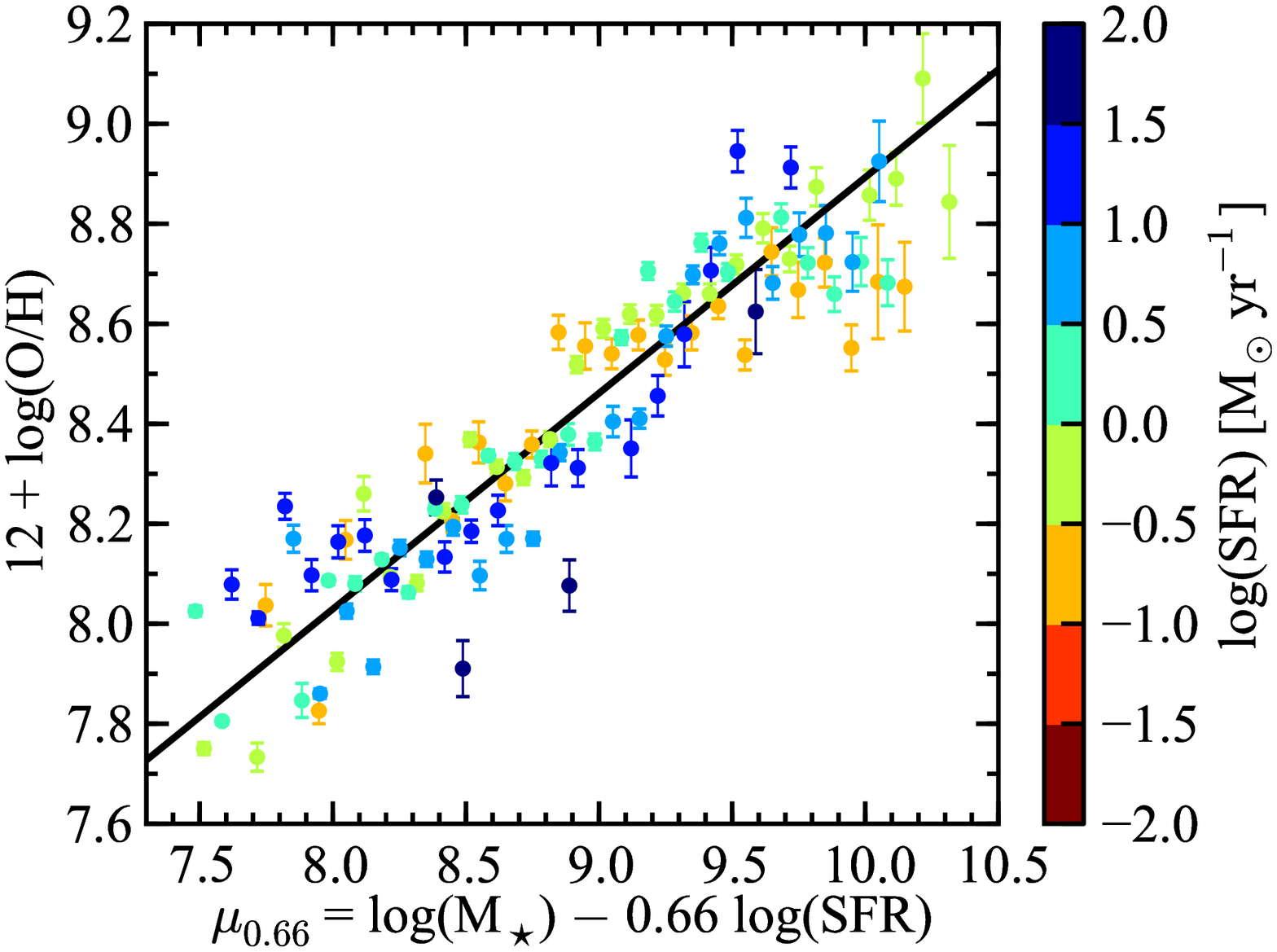}}
\caption{The direct method fundamental metallicity relation for the \mstar--SFR
  stacks (circles color-coded by SFR).  The coefficient (0.66) on log(SFR) in
  the abscissa minimizes the scatter in the FMR (see Equation \ref{eqn:fmr}).
  The black line shows a linear fit to the data, with a slope of 0.43.
\label{fig:fmr}}
\end{figure}

The orientation of the \mstar--$Z$--SFR relation captures the importance of SFR
as a second parameter to the MZR \citep{laralopez2010, mannucci2010}.
\citet{mannucci2010} established the convention that the FMR is the projection
of least scatter found by choosing a free parameter $\alpha$ that minimizes the
scatter in the metallicity vs.~$\mu_\alpha
\equiv$~log(\mstar)~$-$~$\alpha$~log(SFR) plane (Equation \ref{eqn:fmr}).
\citet{mannucci2010} found a value of $\alpha$~=~0.32 for a sample of SDSS
galaxies with metallicities determined with the semi-empirical calibration of
\citet{maiolino2008}.  As metallicity estimates are well known to vary
substantially between different methods, the parameter $\alpha$ may also be
different due to potentially different correlations between the inferred
metallicity and the SFR.  For example, \citet{yates2012} used the
\citetalias{tremonti2004} metallicities, rather than those employed by
\citet{mannucci2010}, and found a lower value of $\alpha$~=~0.19.

\begin{deluxetable}{lc}
\tabletypesize{\small}
\tablecaption{Best Fit $\alpha$\label{table:alpha}}
\tablewidth{0pt}
\tablehead{
\colhead{Calibration} &
\colhead{$\alpha$} \\
\colhead{(1)} &
\colhead{(2)}
}
\startdata
direct method  &  0.66 \\
KK04           &  0.24 \\
M91            &  0.17 \\
Z94            &  0.25 \\
KD02           &  0.12 \\
D02            &  0.34 \\
PP04 N2        &  0.30 \\
PP04 O3N2      &  0.32 \\
\enddata

\tablecomments{Column (1): Metallicity calibration (see Section
\ref{sec:sl_abund} for a more detailed description of the strong line
calibrations).  Column (2): The coefficient on log(SFR) in Equation
(\ref{eqn:fmr}) that minimizes the scatter in the fundamental metallicity
relation.}

\end{deluxetable}

Figure \ref{fig:fmr} shows the fundamental metallicity relation for the
\mstar--SFR stacks (circles color-coded by SFR).  The scatter in metallicity at
fixed $\mu_\alpha$ is minimized for $\alpha$~=~0.66, which is significantly
larger than the $\alpha$ values found by \citet{mannucci2010} and
\citet{yates2012} for metallicities estimated with strong line calibrations.
The scatter for the stacks differs from the scatter for individual galaxies
(like the \citealt{mannucci2010} and \citealt{yates2012} studies) because the
number of galaxies per stack varies.  For a direct comparison, we computed the
value of $\alpha$ for the metallicities derived from various empirical,
semi-empirical, and theoretical strong line calibrations for the stacks with
log(SFR)~>~$-$1.0 (the same SFR range as the stacks with direct method
metallicities) and find low $\alpha$ values ($\alpha$~=~0.12--0.34) that are
consistent with the \citet{mannucci2010} and \citet{yates2012} $\alpha$ values
(see Table \ref{table:alpha}).  The significant difference in $\alpha$ between
the direct method and the strong line methods indicates that the calibrations of
all of the strong line methods have some dependence on physical properties that
correlate with SFR.

The scatter in the direct method FMR ($\sigma$~=~0.13~dex; Figure
\ref{fig:fmr}) is almost a factor of two smaller than the scatter for the
\mstar--SFR stacks with direct method metallicities in the mass--metallicity
plane ($\sigma$~=~0.22~dex; Figure \ref{fig:mzr_sfr}).  This decrease is due to
two features of the \mstar--SFR stacks at fixed SFR shown as the solid colored
lines in Figure \ref{fig:mzr_sfr}: (1) they are substantially offset from one
another; (2) they have similar slopes with minimal overlap.  The former
reflects a strong SFR-dependence on the MZR; the latter corresponds to a
monotonic SFR--metallicity relation at fixed stellar mass.

\begin{figure*}
\centerline{\includegraphics[width=19cm]{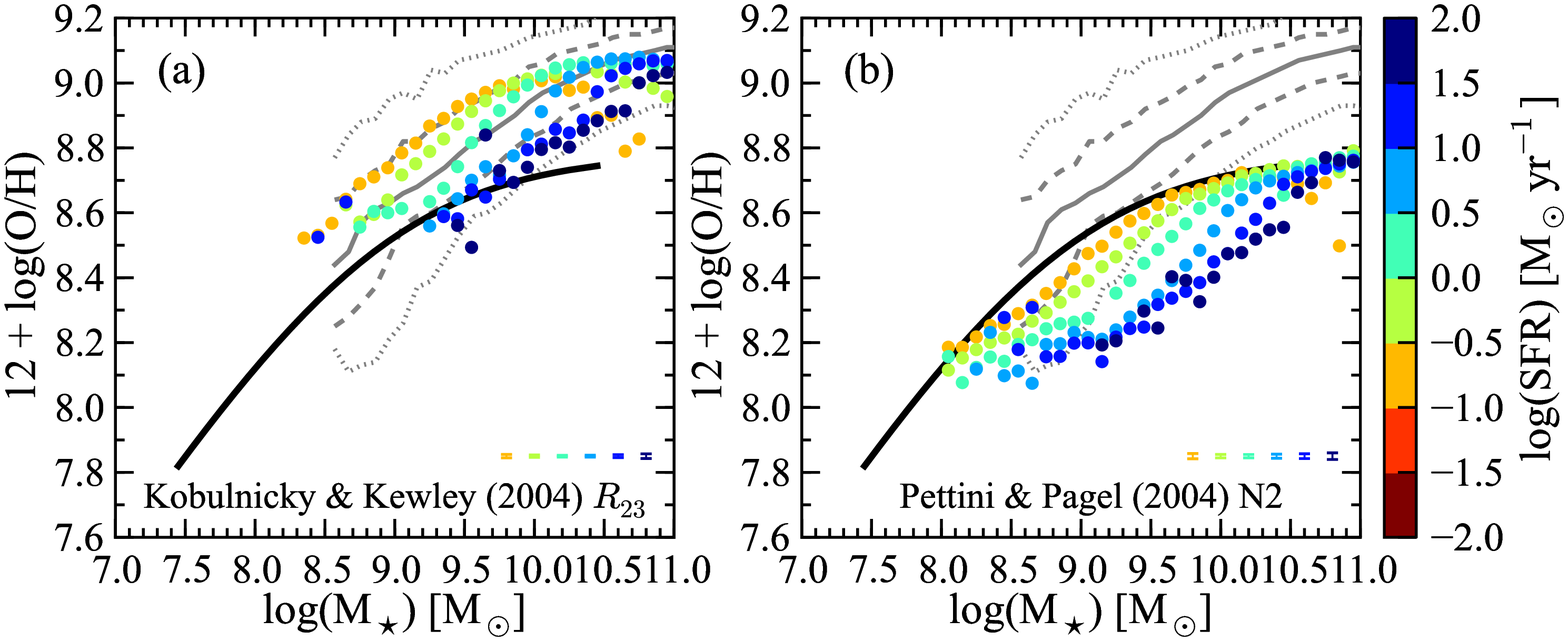}}
\caption{The \mstar--SFR stacks (circles color-coded by SFR) in the
  mass--metallicity plane with metallicities determined with the
  \citet{kobulnicky2004} \rtt\ calibration (panel a) and the \citet{pettini2004}
  N2 calibration (panel b).  The thick black line shows the direct method MZR
  from Figure \ref{fig:mzr}.  The solid, dashed, and dotted gray lines show the
  median, 68\% contour, and 95\% contour, respectively, of the
  \citet{tremonti2004} MZR.  The error bars correspond to the mean error for the
  \mstar--SFR stacks of a given SFR.
\label{fig:mzr_sfr2}}
\end{figure*}

Figure \ref{fig:mzr_sfr2} shows the \mstar--SFR stacks (circles color-coded by
SFR) in the mass--metallicity plane with metallicities determined with two
representative strong line calibrations: the \citet{kobulnicky2004} theoretical
\rtt\ calibration (panel a) and the \citet{pettini2004} empirical N2 calibration
(panel b).  Only stacks with log(\mstar)~$\geq$~8.0 were included in Figure
\ref{fig:mzr_sfr2} because some stacks at lower stellar masses had unphysically
high strong line metallicities; to facilitate comparison with Figure
\ref{fig:mzr_sfr}, only stacks with log(SFR)~>~$-$1.0 are shown in Figure
\ref{fig:mzr_sfr2}.  The stacks in panel (a) show the metallicity from the upper
branch of \rtt, which were selected to have log(\niir/\halpha)~>~$-$1.1
\citep{kewley2008}.  Panel (b) shows stacks with
$-$2.5~<~log(\niir/\halpha)~<~$-$0.3, the calibrated range for the
\citet{pettini2004} N2 calibration according to \citet{kewley2008}.  For
reference, the thick black line shows the direct method MZR.  The median, 68\%
contour, and 95\% contour of the \citetalias{tremonti2004} MZR are indicated by
the solid, dashed, and dotted gray lines, respectively.

The scatter in metallicity about the best fit relation decreases only
marginally from the MZR to the FMR when strong line calibrations are used to
estimate metallicity.  For the \citetalias{kobulnicky2004} and
\citetalias{pettini2004} N2 metallicities, the scatter is reduced by
$\sigma$~=~0.10$\rightarrow$0.09~dex and $\sigma$~=~0.10$\rightarrow$0.07~dex,
respectively.  Figure \ref{fig:mzr_sfr2} also shows that the constant SFR
tracks for the strong line calibrations in the mass--metallicity plane are both
more closely packed and overlap more than those of the direct method.  Figure
\ref{fig:mzr_sfr2} only shows the \mstar--SFR stacks with metallicities from
two strong line calibrations, one theoretical and one empirical, but the minor
reduction in scatter, small spread, and considerable overlap are generic
features of strong line metallicities (the normalization is not).

A qualitative measure of the spread is the difference between the metallicity of
the SFR$_{-0.5}^{0.0}$ (light green) and the SFR$_{0.5}^{1.0}$ (light blue)
stacks at a given stellar mass.  There are 17 stellar mass bins with direct
method metallicities for stacks with these SFRs.  The median metallicity
difference for these pairs of stacks was 0.38 dex for the direct method, 0.15
dex for the \citet{kobulnicky2004} calibration, and 0.13 dex for the
\citet{pettini2004} calibration.  The factor of $\sim$2--3 difference between
the direct method and strong line metallicities translates into an analogous
difference in the SFR-dependence of the MZR.

Another feature of the strong line MZRs at fixed SFR is that different SFR
tracks turn over at different stellar masses.  Low SFR tracks turn over at lower
stellar masses than high SFR tracks, so the sign of the dependence of the MZR on
SFR changes with stellar mass.  At low stellar masses, higher SFR stacks have
lower metallicities; at high stellar masses, the opposite is true---higher SFR
stacks have higher metallicities.  \citet{yates2012} found a similar, but more
dramatic, result for their sample of galaxies that used
\citetalias{tremonti2004} metallicities.  The origin of the weak SFR-dependence
and non-monotonic relation for the strong line calibrations is not obvious, but
we discuss several potentially relevant effects in Section \ref{sec:alpha}.


\section{N/O Abundance}
\label{sec:no}

\begin{figure}
\centerline{\includegraphics[width=9cm]{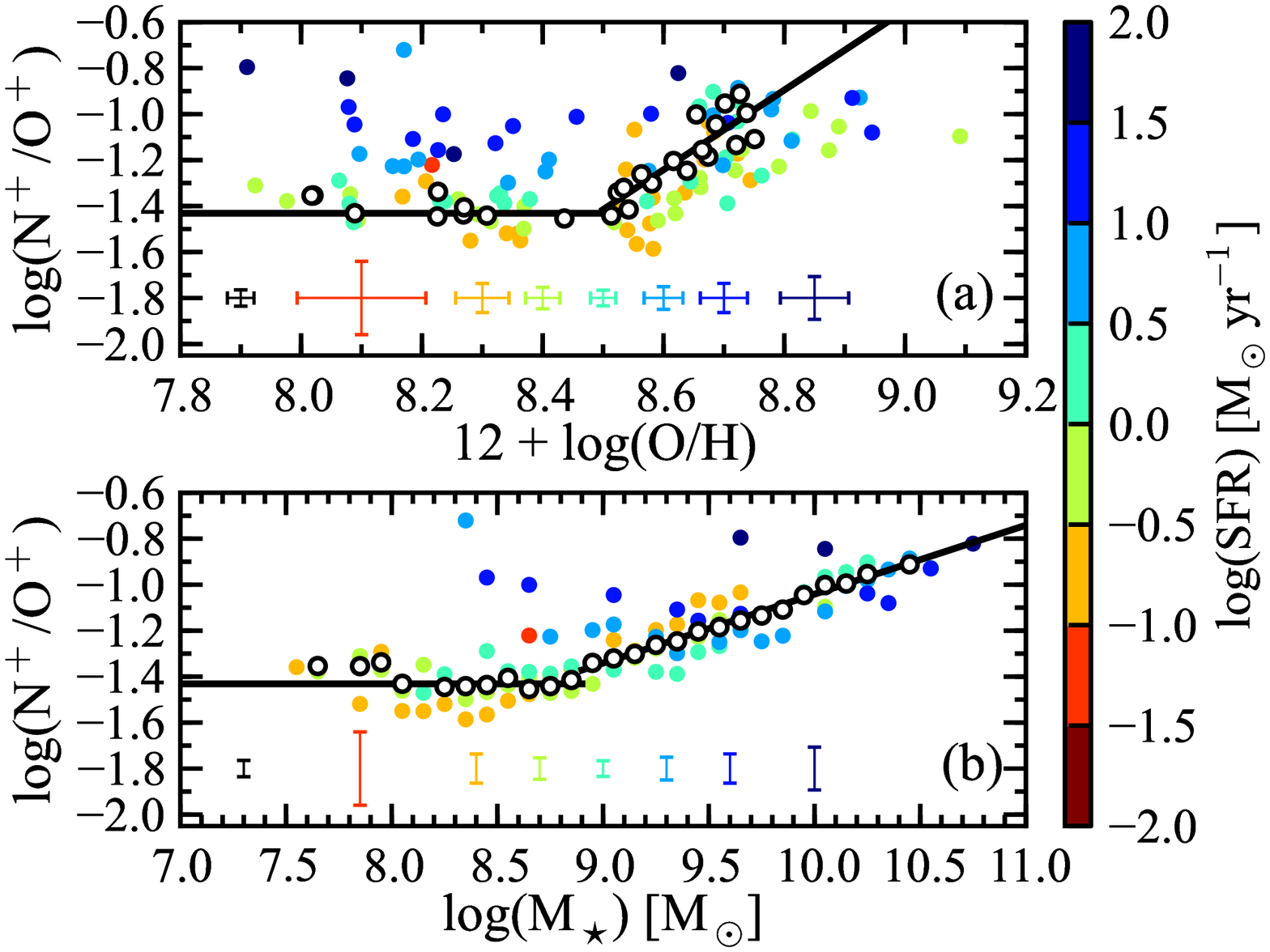}}
\caption{N$^+$/O$^+$ ratio as a function of direct method oxygen abundance
  (panel a) and \mstar\ (panel b) for the \mstar\ stacks (open circles) and
  \mstar--SFR stacks (circles color-coded by SFR).  The horizontal lines show
  the median of the low oxygen abundance (12~+~log(O/H)~<~8.5) and low stellar
  mass (log[\mstar]~<~8.9) data.  The positively sloped lines in panels (a) and
  (b) are linear fits to the stacks with 12~+~log(O/H)~>~8.5 and
  log(\mstar)~>~8.9, respectively, whose fit parameters are given in Table
  \ref{table:no}.  The error bars show the mean error for the \mstar\ stacks
  (black) and each SFR bin of the \mstar--SFR stacks (color-coded by SFR).  If
  N$^+$/O$^+$ is assumed to trace N/O, as is often done, then our results can be
  compared directly to literature results on N/O.  The N/O abundances of the
  stacks are reported in Table \ref{table:metallicity}.
\label{fig:no}}
\end{figure}

Nitrogen provides interesting constraints on chemical evolution because it is
both a primary and secondary nucleosynthetic product.  The yields of primary
elements are independent of the initial metal content of a star but the yields
of secondary elements are not.  In a low metallicity star, the majority of the
seed carbon and oxygen nuclei that will form nitrogen during the CNO cycle are
created during helium burning in the star, so the nitrogen yield of such a star
will scale roughly with the carbon and oxygen yields.  In this case, carbon,
nitrogen, and oxygen all behave like primary elements.  After the ISM becomes
sufficiently enriched, the nitrogen yield of a star principally depends on the
amount of carbon and oxygen incorporated in the star at birth.  The carbon and
oxygen still behave like primary elements, but nitrogen is a secondary
nucleosynthetic product.  Observational studies \citep{vilacostas1993,
vanzee2006a, berg2012} have found clear evidence for primary and secondary
nitrogen at low and high metallicity.  \citet{vilacostas1993} created a simple,
closed box chemical evolution model that quantified the regimes where nitrogen
is expected to behave like a primary and secondary element.  However, modeling
nitrogen enrichment is difficult because of the large uncertainties in stellar
yields and the delay time for nitrogen enrichment relative to oxygen.  Galactic
winds also complicate nitrogen enrichment because they may preferentially eject
oxygen relative to nitrogen \citep{vanzee2006a}.  This is because oxygen is
formed quickly in massive stars and is available to be ejected from galaxies by
winds associated with intense bursts of star formation.  By contrast, the
>100~Myr delay before the release of nitrogen from intermediate mass AGB stars
might be sufficient to protect it from ejections by galactic winds.

In principle, the N/O abundance as a function of oxygen abundance can be used
to disentangle the effects of nucleosynthesis, galactic inflows and outflows,
and different star formation histories on the relative enrichment of nitrogen.
The total N/O ratio is a difficult quantity to measure because \niiis\ lines
are not readily observable, so N$^+$/O$^+$ is used frequently as a proxy for
N/O.  This assumption is supported by the photoionization models of
\citet{garnett1990}, which showed that the ionization correction factor from
N$^+$/O$^+$ to N/O should be $\sim$1 to within 20\%.  Because the ionization
factor should be close to unity, most papers in the literature
\citep[e.g.,][]{vilacostas1993} that show N/O have assumed N/O~=~N$^+$/O$^+$.
For transparency, we plot N$^+$/O$^+$ as a function of direct method oxygen
abundance in Figure \ref{fig:no}a for the \mstar\ and \mstar--SFR stacks.  We
measured the ionic abundances of N$^+$ and O$^+$ with the direct method under
the assumption that \teoii\ represents \tii\ (see Section \ref{sec:abund}).

\begin{deluxetable}{cccccc}
\tabletypesize{\small}
\tablecaption{N/O vs. O/H and \mstar\ Fit Parameters for \mstar\ Stacks
\label{table:no}}
\tablewidth{0pt}
\tablehead{
\colhead{Abscissa} &
\colhead{Slope} &
\colhead{y-intercept} &
\colhead{Dispersion} &
\colhead{Fit Range} \\
\colhead{(1)} &
\colhead{(2)} &
\colhead{(3)} &
\colhead{(4)} &
\colhead{(5)}
}
\startdata
12~+~log(O/H)  &  0  &  $-$1.43  &  0.04  &  12~+~log(O/H)~<~8.5 \\
12~+~log(O/H)  &  1.73  &  $-$16.15  &  0.08  & 12~+~log(O/H)~>~8.5 \\
\mstar\  &  0  &  $-$1.43  &  0.04  & \mstar~<~8.9 \\
\mstar\  &  0.30  &  $-$4.04  &  0.01  & \mstar~>~8.9 \\
\enddata

\tablecomments{Column (1): N/O as a function of 12~+~log(O/H) or \mstar.  Column
  (2): Slope of linear fit (set to 0 for first and third rows).  Column (3):
  y-intercept of linear fit.  Column (4): Dispersion around fit.  Column (5):
  Range in 12~+~log(O/H) or \mstar\ of the fit.}

\end{deluxetable}

At low metallicity (12~+~log(O/H)~<~8.5), we find that the \mstar\ stacks have
an approximately constant value of N$^+$/O$^+$, which is expected for primary
nitrogen.  These stacks have a median of log(N$^+$/O$^+$)~=~$-$1.43 (indicated
by the horizontal line\footnote{We do not show a fit to these points because of
the strong \textit{a priori} expectation of a constant N$^+$/O$^+$ at low
metallicity (and low mass); however, a linear fit would have a slope of $-$0.21.
The analogous slope for the low mass N$^+$/O$^+$--\mstar\ relation is
$-$0.08.}), which is consistent with other studies of \hii\ regions and dwarf
galaxies \citep{vilacostas1993}.  At 12~+~log(O/H)~=~8.5, there is a sharp
transition where N$^+$/O$^+$ increases steeply with oxygen abundance
(slope~=~1.73), which shows that nitrogen is acting like a secondary element.
Previous observations \citep[e.g.,][]{vilacostas1993} have found a smoother
transition between primary and secondary nitrogen and a shallower slope in the
secondary nitrogen regime, albeit with large dispersion that could be obscuring
these features.  The fit parameters of the N$^+$/O$^+$--O/H relation for the
\mstar\ stacks is presented in Table \ref{table:no}.

The \mstar\ stacks form a tight sequence with a dispersion of only
$\sigma$~=~0.08~dex, compared to a more typical dispersion of
$\sigma$~$\sim$~0.3~dex for individual objects (e.g., \citealt{henry2000}).  A
plausible explanation for the additional scatter in the N/O--O/H relation for
individual galaxies is the time-dependence of N/O caused by the difference in
enrichment timescales of oxygen and nitrogen following a burst of star
formation.  The \mstar--SFR stacks show a larger dispersion than the \mstar\
stacks, potentially because these stacks contain fewer galaxies.  The low and
moderate SFR stacks (SFR$_{-1.0}^{-0.5}$, SFR$_{-0.5}^{0.0}$, and
SFR$_{0.0}^{0.5}$) follow the general trend of the \mstar\ stacks; however, the
high SFR stacks (SFR$_{0.5}^{1.0}$, SFR$_{1.0}^{1.5}$, and SFR$_{1.5}^{2.0}$)
have higher N$^+$/O$^+$ at a given oxygen abundance, which may be because these
galaxies have experienced a large inflow of gas that would lower O/H at fixed
N/O (i.e., move galaxies to the left in Figure \ref{fig:no}a).  Another
consequence of a vigorous burst of star formation is the production of
Wolf-Rayet stars that can enrich the gas in nitrogen for a brief period before
the oxygen enrichment from the subsequent SNe II \citep{berg2011}.  We see
evidence for Wolf-Rayet features, such as \heii, in some of our stacks,
especially at low mass.

Some of the features in the N$^+$/O$^+$--O/H relation are clarified by the
associated relation between N$^+$/O$^+$ and stellar mass, which is shown in
Figure \ref{fig:no}b for the \mstar\ and \mstar--SFR stacks (see Table
\ref{table:no} for the fit parameters of the N$^+$/O$^+$--\mstar\ relation for
the \mstar\ stacks).  Similar to Figure \ref{fig:no}a, there is a primary
nitrogen plateau in N$^+$/O$^+$ at low stellar mass (log[\mstar]~<~8.9) and a
steady increase in N$^+$/O$^+$ due to secondary nitrogen enrichment above
log(\mstar)~=~8.9 (slope~=~0.30).  However, in the secondary nitrogen regime,
the N$^+$/O$^+$--\mstar\ relation has a much lower dispersion
($\sigma$~=~0.01~dex) than the N$^+$/O$^+$--O/H relation ($\sigma$~=~0.08~dex).
Some of the decreased dispersion is due to the larger dynamic range of stellar
mass relative to oxygen abundance, but the tightness of the
N$^+$/O$^+$--\mstar\ relation suggests that the enrichment of nitrogen relative
to oxygen is well-behaved on average.  The essentially zero intrinsic
dispersion in the N$^+$/O$^+$--\mstar\ relation can be used to quantify the
effect of gas inflow and galactic winds on enrichment if all of the scatter in
N$^+$/O$^+$ at a given O/H is due to gas flows into and out of galaxies.  As in
the N$^+$/O$^+$--O/H relation, the low and moderate SFR stacks
(SFR$_{-1.0}^{-0.5}$, SFR$_{-0.5}^{0.0}$, and SFR$_{0.0}^{0.5}$) roughly
coincide with the \mstar\ stacks.  The high SFR stacks (SFR$_{0.5}^{1.0}$,
SFR$_{1.0}^{1.5}$, and SFR$_{1.5}^{2.0}$) still tend to be more nitrogen
enriched than the \mstar\ stacks at a fixed \mstar, but the discrepancy has
decreased.  The N$^+$/O$^+$--\mstar\ diagram is less sensitive to dilution
(traced by SFR) because the high SFR galaxies with low O/H are less significant
outliers when shown as a function of stellar mass.

The N/O--\mstar\ relation has been previously investigated by
\citet{perezmontero2009b} and \citet{perezmontero2013}, who used strong line
methods to estimate N/O.  They found that N/O increased steadily with stellar
mass and did not show a plateau at low stellar mass associated with primary
nitrogen enrichment, in contrast to the direct method N/O--\mstar\ relation.
However, \citet{perezmontero2013} showed that the strong line N/O--\mstar\
relation is nearly independent of SFR, which is roughly consistent with our
finding that the N/O--\mstar\ relation has only a mild dependence on SFR,
particularly at log(\mstar)~$\gtrsim$~9.0.

\section{Discussion}
\label{sec:discussion}

\subsection{Comparison to a Previous Analysis That Used Auroral Lines from Stacked Spectra}

\begin{figure}
\centerline{\includegraphics[width=9cm]{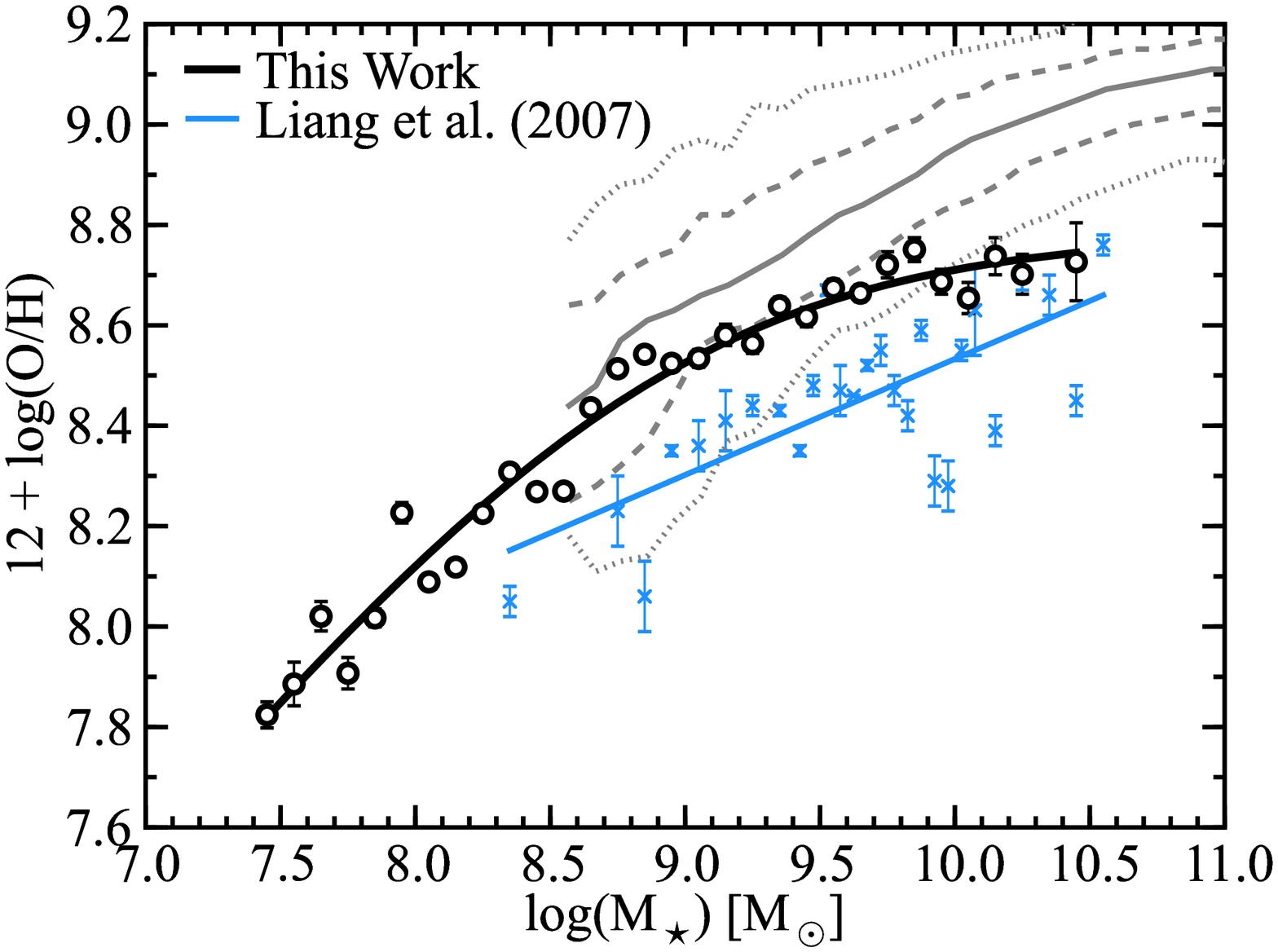}}
\caption{Our direct method MZR (open circles and thick black line) and the
  \citet{liang2007} direct method MZR (blue crosses and line).  For reference,
  the solid, dashed, and dotted gray lines show the median, 68\% contour, and
  95\% contour, respectively, of the \citet{tremonti2004} MZR.
\label{fig:liang}}
\end{figure}

\citet{liang2007} stacked SDSS spectra and applied the direct method to estimate
the MZR, although their study differs from ours in a number of important
respects. First, their study is based on DR4 spectroscopy of 23,608 galaxies,
which is approximately an order of magnitude fewer than our sample. Second, they
implemented a minimum \oii\ EW criterion to select the input galaxies to their
stacks in order to increase the SNR of their stacked spectra. Finally, they only
measured \teoii\ from the \oiia\ lines and then inferred \teoiii\ (and the
O$^{++}$ ionic abundance) from the \tii--\tiii\ relation provided by
\citet{izotov2006}. These differences are likely responsible for the offset
between our MZR and the \citet{liang2007} MZR, the absence of a turnover in
their MZR, and their greater scatter as shown in Figure \ref{fig:liang}.

The \oiis\ selection criterion can readily explain part of the offset between
our MZRs. \citet{liang2007} only selected galaxies with above average \oii\ EW
(at fixed mass) for galaxies with log(\mstar)~<~10 and required a more stringent
EW(\oiis)~>~30~\AA\ for galaxies with log(\mstar)~>~10. As a result of this
selection, their stacks have systematically higher SFRs by approximately 0.15 to
0.2 dex. This in turn biases the stacks to lower metallicities because of the
\mstar--$Z$--SFR relation \citep{mannucci2010, laralopez2010}.  The magnitude of
this effect ($\sim$0.05--0.08 dex) accounts for part of the difference between
the MZRs. Another effect of this selection is that the increase in average SFR
increases the turnover mass and makes it less distinct (see Figure
\ref{fig:mzr_sfr}).

The turnover mass is also not apparent in their MZR due to the greater scatter,
which is largely due to their order of magnitude smaller sample. The scatter
around the linear fit from log(\mstar)~=~8.0--10.5 for their data is
$\sigma$~=~0.12~dex. The scatter around an asymptotic logarithmic fit (Equation
\ref{eqn:alog}) is reduced only to $\sigma$~=~0.11~dex.  An asymptotic
logarithmic fit has an additional degree of freedom relative to a linear fit,
so the marginal improvement in $\sigma$ suggests that the \citet{liang2007} MZR
can be sufficiently characterized by a linear fit.  Over the same mass range,
the scatter around the asymptotic logarithmic fit of our data (thick black
line) is only $\sigma$~=~0.03~dex, or a factor of four smaller. The smaller
scatter in our MZR enables a clear identification of the turnover.

The method employed by \citet{liang2007} to estimate the oxygen abundance is
also distinct from ours and may explain the rest of the discrepancy in the
normalization difference between our studies. The \citet{liang2007} study relies
solely on the \oiia\ auroral lines to measure \teoii, which is used to infer
\teoiii\ and the O$^{++}$ abundance by applying the \tii--\tiii\ relation and
\teoiii--(O$^{++}$/H$^+$) formula from \citet{izotov2006}.  They did not detect
\oiiia\ in their stacks, which they only binned in stellar mass, because they
had fewer galaxies per bin. The stellar continuum subtraction may also have
affected the detection of \oiiia\ because of its proximity to the \hgamma\
stellar absorption feature, whereas the stellar continuum is comparatively
featureless in the vicinity of the \oiia\ lines. \citet{liang2007} used the
\citet{bruzual2003} spectral templates, rather than the empirical and higher
resolution {\sc miles} templates that we have adopted (see Section
\ref{sec:scs}), and this difference may also have played an important role. As a
consequence of their lack of a detection of \oiiia, their oxygen abundance
estimate depends on the quality of the assumption that the galaxies obey the
\tii--\tiii\ relation of \citet{izotov2006}. Our empirical measurements of \tii\
and \tiii\ indicate that this assumption underestimates \tiii\ and overestimates
O$^{++}$/H, which may partly explain why our MZRs are in better agreement at
high mass where O$^+$ is the dominant ionization state of oxygen.

\subsection{Temperature and Metallicity Discrepancies}
Temperatures and metallicities of \hii\ regions measured with the direct method
do not always agree with those measured with other techniques.  For example,
temperatures measured with the direct method tend to be systematically higher
than those measured from the Balmer continuum \citep{peimbert1967}.  Also, the
metallicities determined from optical recombination lines (e.g., \ciiorl\ and
\oiiorl) and far-IR fine-structure lines (e.g., \oiiis~52,~88~$\mu$m) tend to be
0.2--0.3 dex higher than those from collisionally excited lines
\citep{garciarojas2007, bresolin2008, esteban2009}.  The exact cause of these
temperature and abundance discrepancies is currently not understood.

\citet{peimbert1967} proposed that temperature fluctuations and gradients in
\hii\ regions cause direct method temperatures to be systematically
overestimated, while direct method metallicities are underestimated.  To
account for temperature variations across a nebula, he introduced the concept
of $t^2$, the root mean square deviation of the temperature from the mean.
Estimating $t^2$ has proven to be difficult, so most direct method metallicity
studies assume $t^2$~=~0.  However, optical recombination lines and far-IR
fine-structure lines \citep{garnett2004a} are less sensitive to temperature
than collisionally excited lines, so they could be used to estimate $t^2$ if
the discrepancy between the metallicity determined from collisionally excited
lines and optical recombination lines or far-IR fine-structure lines is assumed
to be caused by temperature fluctuations.  The few studies that have measured
optical recombination lines (e.g., \citealt{garciarojas2007, esteban2009}) find
that values of $t^2$~=~0.03--0.07 are necessary to increase the direct method
metallicities by 0.2--0.3 dex to match the optical recombination line
metallicities.

Recently, \citet{nicholls2012} suggested that the electron energy distribution
could be the cause of the temperature and metallicity discrepancies.
Specifically, they questioned the widespread assumption that the electrons are
in thermal equilibrium and can be described by a Maxwell-Boltzmann distribution.
Instead, they suggested that a there might be an excess of high energy electrons
and proposed that a $\kappa$-distribution is a more appropriate description of
the electron energy distribution.  The $\kappa$-distribution is based on direct
measurements of solar system plasmas.  Assuming a $\kappa$-distribution for an
\hii\ region lowers the derived temperature, increases the inferred metallicity,
and could potentially resolve the discrepancy between the temperatures and
metallicities found with optical recombination lines and collisionally excited
lines.

Models of \hii\ regions by \citet{stasinska2005} indicate that metallicities
based on the direct method could suffer from systematic biases in metal-rich
\hii\ regions.  She finds that measuring metallicity from \teoiii\ and \tenii\
tends to dramatically underestimate the true metallicity for 12~+~log(O/H)~>~8.6
(see her Figure 1).  The situation does not improve if metallicities are
computed with only \tenii\ because the derived metallicity can wildly
overestimate or underestimate the true metallicity depending on the physical
conditions and geometry of the \hii\ region.  However, there are two key
differences between the models of \citet{stasinska2005} and the measurements
made in this study that could minimize the bias.  First, we measured the
temperature of the low ionization region from \teoii, not \tenii.  Second, we
analyzed spectra of galaxy stacks and not individual \hii\ regions, which may
average out the large predicted errors.  Our stacks generally increase smoothly
in metallicity as a function of \mstar, which does not rule out systematic error
but minimizes the impact of the individual, catastrophic errors highlighted in
\citet{stasinska2005}.

\subsection{Strong Line Calibrations and the SFR-dependence of the FMR}
\label{sec:alpha}
\begin{figure*}
\centerline{\includegraphics[width=19cm]{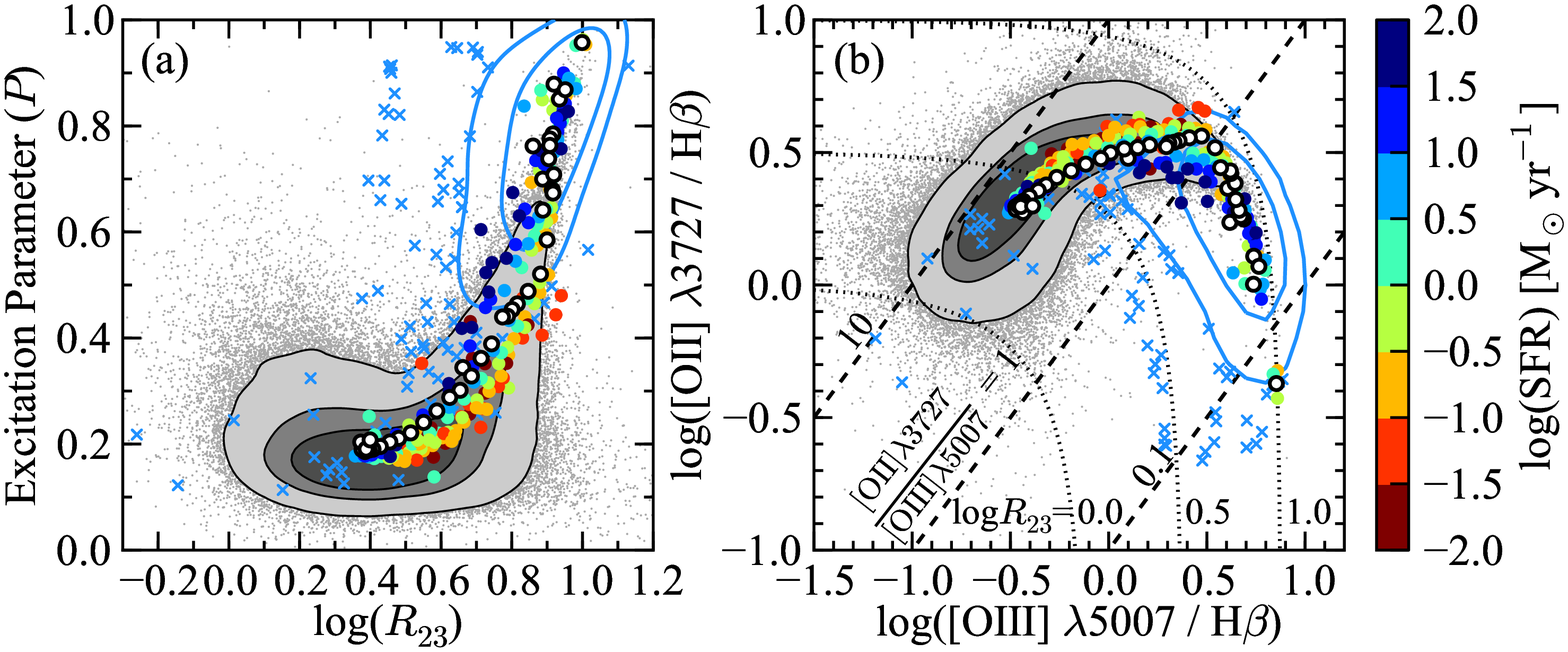}}
\caption{Panel (a) shows the excitation parameter, $P =$ \oiii\ / (\oii\ +
  \oiii), as a function of \rtt\ = (\oii\ + \oiii) / \hbeta.  Panel (b) shows
  log(\oii~/~\hbeta) versus log(\oiiir~/~\hbeta).  The gray scale contours
  (50\%, 75\%, and 95\%) and gray points correspond to SDSS star-forming
  galaxies.  The white and colored circles represent the \mstar\ and \mstar--SFR
  (color-coded by SFR) stacks, respectively.  The light blue contours (50\% and
  75\%) and light blue crosses show \hii\ regions with direct method
  metallicities from the \citet{pilyugin2012} compilation.  In panel (b), the
  dashed and dotted lines show lines of constant \oii~/~\oiiir\ and \rtt,
  respectively.  The stacks trace the overall galaxy distribution better than
  \hii\ regions, especially at lower excitation parameters.  The \hii\ regions
  tend to have high excitation parameters because the auroral line flux is a
  strong function of metallicity and hence \rtt.
\label{fig:PR23_O2O3}}
\end{figure*}

The MZRs and FMRs based on strong line calibrations (Figure \ref{fig:mzr_sfr2})
have a much weaker dependence on SFR than the direct method MZR and FMR (Figures
\ref{fig:mzr}--\ref{fig:fmr}).  Relative to the direct method MZR and FMR, the
strong line MZR and FMR have (1) a smaller spread in the mass--metallicity plane
(compare Figures \ref{fig:mzr_sfr} and \ref{fig:mzr_sfr2}), (2) a smaller
reduction in scatter from the MZR to the FMR (see Section \ref{sec:fmr}), and
(3) a smaller value of $\alpha$ (see Table \ref{table:alpha}).  This trend is a
generic feature of strong line calibrations that holds for both empirical and
theoretical calibrations and for all strong line indicators (\rtt, N2, N2O2, and
O3N2) that we used.  Since a strong line calibration is only applicable to the
physical conditions spanned by the calibration sample or model, it is important
to understand the physical properties of the calibration sample for empirical
calibrations and the assumptions behind the \hii\ region models that underlie
theoretical calibrations.

Figure \ref{fig:PR23_O2O3} compares excitation parameter ($P$) and \rtt\ (panel
a) and \oiis\ and \oiiis\ fluxes relative to \hbeta\ (panel b) for galaxies,
stacks of galaxies, and \hii\ regions.  The gray contours (50\%, 75\%, and 95\%)
and points indicate SDSS star-forming galaxies, whose line flux measurements
come from the MPA-JHU catalog \citep{tremonti2004}, after we corrected their
measured values for intrinsic reddening.  The stacks are shown by the open and
colored circles.  Extragalactic \hii\ regions with direct method metallicities
are represented by the light blue contours and crosses.  The dereddened line
fluxes of the \hii\ regions come from the literature compilation by
\citet{pilyugin2012}\footnote{The original data can be found in
\citet{bresolin2004}, \citet{bresolin2005}, \citet{bresolin2007},
\citet{bresolin2009a}, \citet{bresolin2009b}, \citet{campbell1986},
\citet{castellanos2002}, \citet{deblok1998}, \citet{esteban2009},
\citet{fierro1986}, \citet{french1980}, \citet{fricke2001}, \citet{garnett1997},
\citet{garnett2004b}, \citet{gonzalezdelgado1994}, \citet{guseva2000},
\citet{guseva2001}, \citet{guseva2003a}, \citet{guseva2003b},
\citet{guseva2004}, \citet{guseva2009}, \citet{guseva2011}, \citet{hagele2008},
\citet{hawley1978}, \citet{hodge1995}, \citet{izotov1994}, \citet{izotov1997},
\citet{izotov1998a}, \citet{izotov1998b}, \citet{izotov1999},
\citet{izotov2001}, \citet{izotov2004a}, \citet{izotov2004b},
\citet{izotov2009}, \citet{izotov2011}, \citet{kehrig2004}, \citet{kehrig2011},
\citet{kennicutt2001}, \citet{kennicutt2003}, \citet{kinkel1994},
\citet{kniazev2000}, \citet{kobulnicky1997a}, \citet{kunth1983},
\citet{kwitter1981}, \citet{lee2003a}, \citet{lee2003b}, \citet{lee2004},
\citet{lee2005}, \citet{lequeux1979}, \citet{lopezsanchez2004},
\citet{lopezsanchez2007}, \citet{lopezsanchez2009}, \citet{lopezsanchez2011},
\citet{luridiana2002}, \citet{magrini2009}, \citet{mccall1985},
\citet{melbourne2004}, \citet{melnick1992}, \citet{miller1996},
\citet{noeske2000}, \citet{pagel1980}, \citet{pagel1992}, \citet{pastoriza1993},
\citet{peimbert1986}, \citet{pena2007}, \citet{perezmontero2009a},
\citet{popescu2000}, \citet{pustilnik2002}, \citet{pustilnik2003a},
\citet{pustilnik2003b}, \citet{pustilnik2005}, \citet{pustilnik2006},
\citet{rayo1982}, \citet{saviane2008}, \citet{sedwick1981},
\citet{skillman1985}, \citet{skillman1993}, \citet{skillman2003},
\citet{stanghellini2010}, \citet{terlevich1991}, \citet{thuan1995},
\citet{thuan1999}, \citet{torrespeimbert1989}, \citet{tullmann2003},
\citet{vanzee1997}, \citet{vanzee1998}, \citet{vanzee2000}, \citet{vanzee2006a},
\citet{vanzee2006b}, \citet{vilchez1988}, \citet{vilchez2003},
\citet{webster1983}, and \citet{zahid2011b}.}.

Figure \ref{fig:PR23_O2O3}a shows the excitation parameter $P$ as a function of
\rtt.  Excitation increases upwards, but \rtt\ is double-valued with
metallicity, so metallicity increases to the left for objects on the upper
branch (the majority of the galaxies and stacks) and increases to the right for
objects on the lower branch (most of the compiled \hii\ regions).  Figure
\ref{fig:PR23_O2O3}b displays another projection of the same data in the space
defined by the dereddened \oii\ and \oiiir\ line fluxes relative to \hbeta.  The
dotted lines show constant \rtt\ values, and the dashed lines mark constant
\oii/\oiiir\ values.

In Figure \ref{fig:PR23_O2O3}, the compiled \hii\ regions predominantly overlap
with the high excitation and high \rtt\ tail of the galaxy distribution in
Figure \ref{fig:PR23_O2O3}a and the analogous high \oiiir\ tail of the galaxy
distribution in Figure \ref{fig:PR23_O2O3}b, which corresponds to low
metallicity galaxies.  The compiled \hii\ regions have direct method
metallicities and therefore at least one detectable auroral line, usually
\oiiia.  Because the strength of the auroral lines, especially \oiiia, is a
strong function of metallicity and excitation parameter, these \hii\ regions
were effectively selected to have low metallicities and high excitation
parameters.  Thus, they are not representative of the typical conditions found
in the \hii\ regions of the galaxy sample.  Empirical calibrations, which are
based on samples of \hii\ regions with direct method metallicities, are not well
constrained in the high metallicity, low excitation regime where most galaxies
and their constituent \hii\ regions lie.  For example, \citet{moustakas2010}
recommended only using the empirical \citet{pilyugin2005} \rtt\ calibration for
objects with $P$~>~0.4.  When empirical calibrations are applied to large galaxy
samples, galaxy metallicities are systematically underestimated, particularly at
low excitation and high metallicity \citep{moustakas2010}.  Similarly, MZRs
based on empirical calibrations may have an artificially weak dependence on SFR.

The typical excitation conditions and \rtt\ values of the stacks are much better
matched to the overall galaxy distribution than the compiled \hii\ regions with
direct method metallicities.  The stacks probe to both lower excitation
($P~\approx~0.2$) and higher metallicity (\rtt~$\approx$~0.4) than the bulk of
the compiled \hii\ regions.  The stacks do not continue to low \rtt\ values
($<$0.3), a region of parameter space populated by the most massive and
metal-rich galaxies in our sample.  The \oii\ and \oiiir\ line fluxes of these
galaxies vary significantly, even at the same stellar mass and SFR.  While the
stacks do not reach the lowest \rtt\ values of the galaxies, they still trace
the average \rtt\ values of the galaxies in each stack.

Theoretical calibrations are based on stellar population synthesis models, like
STARBURST99 \citep{leitherer1999}, and photoionization models, such as MAPPINGS
\citep{sutherland1993, groves2004a, groves2004b} and CLOUDY \citep{ferland1998}.
The stellar population synthesis model generates an ionizing radiation field
that is then processed through the gas by the photoionization model.  The
parameters in the stellar population synthesis model include stellar
metallicity, age of the ionizing source, initial mass function, and star
formation history.  In the photoionization model, the electron density and the
ionization parameter are adjustable parameters.  Because the model grids can
span a wide range of parameter space, particularly in metallicity and excitation
parameter, theoretical calibrations have an advantage over empirical
calibrations at high metallicity and low excitation, where empirical
calibrations are not strictly applicable.

However, metallicities derived with theoretical calibrations can be
significantly higher (up to 0.7 dex; see \citealt{kewley2008}) than direct
method metallicities.  The most likely cause of this offset is the breakdown of
one or more of the assumptions about the physics of \hii\ regions in the stellar
population synthesis or photoionization models.  In the stellar population
synthesis models, the ionizing source is usually treated as a zero age main
sequence starburst, which is not applicable for older star clusters
\citep{berg2011}, and the line fluxes can change appreciably as a cluster (and
the associated \hii\ region) ages.  As elucidated by \citet{kewley2008}, there
are three main issues with the photoionization models.  First, they treat the
nebular geometry as either spherical or plane-parallel, which may not be
appropriate for the true geometries of the \hii\ regions.  Second, the fraction
of metals depleted onto dust grains is poorly constrained by observations (see
\citealt{draine2003, jenkins2009}) but is a required parameter of the
photoionization models. Third, they assume that the density distribution of the
gas and dust as smooth, when it is clumpy.  While all these assumptions might
break down to some degree, it is unknown which assumption or assumptions causes
metallicities based on theoretical strong line calibrations to be offset from
the direct method metallicities, but it is conceivable that the weak SFR
dependence of theoretical strong line calibration MZRs is also due to these
assumptions.

One of the most intriguing findings of the \citet{mannucci2010} and
\citet{laralopez2010} studies is that high redshift observations are consistent
with no redshift evolution of the strong line FMR up to $z = 2.5$ and $z = 3.5$,
respectively.  Given the large discrepancies between the local strong line and
direct method FMRs, a fair comparison between the local direct method FMR and a
high redshift strong line FMR is not possible.  An interesting test would be to
check if high redshift direct method metallicity measurements are consistent
with the local direct method FMR.  A few studies \citep{hoyos2005, kakazu2007,
yuan2009, erb2010, brammer2012} have reported direct method metallicities at
higher redshifts ($z \sim$ 0.7--2.3), but none simultaneously provide the
stellar masses and SFRs of the galaxies.  Since the FMR and its evolution
provide important constraints on theoretical galaxy evolution models and form
the basis of empirical galaxy evolution models \citep{zahid2012b, peeples2013},
future studies that measure all three of these parameters would be valuable.

\subsection{Physical Processes Governing the MZR and \mstar--$Z$--SFR Relation}
\label{sec:physics}

Understanding the baryon cycling of galaxies relies heavily on the adopted
relations between stellar mass, metallicity, and SFR.  Traditionally, the MZR
and \mstar--$Z$--SFR relation have been measured with strong line methods.  In
this study, we have used the more reliable direct method to measure the MZR and
\mstar--$Z$--SFR relation.  The direct method MZR (Figure \ref{fig:mzr}) spans
three orders of magnitude in stellar mass from log(\mstar)~=~7.4--10.5 and thus
simultaneously extends the MZR to lower masses by an order of magnitude compared
to strong line MZRs (e.g., \citetalias{tremonti2004}) and resolves the high mass
turnover.  The features of the direct method MZR that most strongly influence
the physical interpretations are its low mass slope
(O/H~$\propto$~\mstar$^{1/2}$), its turnover mass (log[\mstar]~=~8.9), and its
normalization (12~+~log(O/H)$_\mathrm{asm}$~=~8.8).  The SFR-dependence of the
MZR (see Figures \ref{fig:mzr_sfr} and \ref{fig:fmr}) also serves as an
important observational constraint for galaxy evolution models.  We find that
the MZR depends strongly on SFR ($\alpha$~=~0.66; Figure \ref{fig:fmr}) at all
stellar masses.

The MZR and \mstar--$Z$--SFR relation are shaped by gas inflows, gas outflows,
and star formation.  The interplay between these three processes is complex, so
hydrodynamic galaxy simulations (e.g., \citealt{brooks2007};
\citealt{finlator2008}; \citealt{dave2011a}; \citealt{dave2011b}) and analytic
models (e.g., \citealt{peeples2011}; \citealt{dave2012}) have been used to
establish a framework to interpret the observations in a physical context.
Below we briefly discuss the physical implications of our results within the
formalisms of \citet{peeples2011} and \citet{finlator2008}.

\citet{peeples2011} developed an analytic model for understanding the importance
of outflows in governing the MZR based on the assumption that galaxies follow
zero scatter relations between stellar mass, gas fraction, metallicity, outflow
efficiency, and host halo properties.  In their formalism, the primary variable
controlling the MZR is the metallicity-weighted mass-loading parameter,
\begin{equation}
\zeta_\mathrm{wind} \equiv \left(\frac{Z_\mathrm{wind}}{Z_\mathrm{ISM}}\right)
\left(\frac{\dot{M}_\mathrm{wind}}{\dot{M}_\star}\right),
\label{eqn:zeta1}
\end{equation}
where $Z_\mathrm{wind}$ and $Z_\mathrm{ISM}$ are the wind and ISM metallicities,
respectively, and $\dot{M}_\mathrm{wind}/\dot{M}_\star$ is the unweighted
mass-loading parameter.  $\zeta_\mathrm{wind}$ can be expressed in terms of the
MZR and the stellar mass--gas fraction relation by rearranging their
Equation~(20):
\begin{equation}
\zeta_\mathrm{wind} = y/Z_\mathrm{ISM} - 1 - \alpha F_\mathrm{gas},
\label{eqn:zeta2}
\end{equation}
where $y$ is the nucleosynthetic yield, $\alpha$ is a parameter of order unity
(see their Equation 11), and $F_\mathrm{gas} \equiv M_\mathrm{gas} / M_\star$ is
the gas fraction.

If we adopt the \citet{peeples2011} formalism and their fiducial yield and
stellar mass--gas fraction relation, then we can solve for the
\mstar--$\zeta_\mathrm{wind}$ relation implied by the direct method MZR.  This
direct method \mstar--$\zeta_\mathrm{wind}$ relation starts at high
$\zeta_\mathrm{wind}$ ($\zeta_\mathrm{wind}$~$\sim$~15) for low mass galaxies
(log[\mstar]~=~7.5).  Then, $\zeta_\mathrm{wind}$ decreases with increasing
stellar mass, eventually flattening and approaching a constant
$\zeta_\mathrm{wind}$ ($\zeta_\mathrm{wind}$~$\sim$~2) above the turnover mass
(log[\mstar]~=~8.9).  Since the \citetalias{denicolo2002} MZR has a similar
shape and normalization to the direct method MZR from log(\mstar)~=~8.5--10.5,
the direct method \mstar--$\zeta_\mathrm{wind}$ relation resembles the
\citetalias{denicolo2002} \mstar--$\zeta_\mathrm{wind}$ relation shown in Figure
6 of \citet{peeples2011}.  Also, the direct method MZR implies a similar
behavior for $Z_\mathrm{wind}$ and $Z_\mathrm{ISM}$ as a function of stellar
mass as the \citetalias{denicolo2002} MZR (see their Figure 9).  The ratio of
$Z_\mathrm{wind}/Z_\mathrm{ISM}$ inversely correlates with how efficiently winds
entrain ambient ISM.  If we adopt the simple relation between metallicity and
the unweighted mass-loading parameter from \citet{finlator2008}, $Z_\mathrm{ISM}
\approx y / (1 + \dot{M}_\mathrm{wind}/\dot{M}_\star)$, then the direct method
MZR implies an efficiency of mass ejection that scales as
$\dot{M}_\mathrm{wind}/\dot{M}_\star \propto M_\star^{-1/2}$ for
log(\mstar)~$\lesssim$~9.0.  The higher $\zeta_\mathrm{wind}$ for low mass
galaxies relative to high mass galaxies could be due to more enriched winds
(larger $Z_\mathrm{wind}/Z_\mathrm{ISM}$) or more efficient mass ejection by
winds (larger $\dot{M}_\mathrm{wind}/\dot{M}_\star$) or both.
\citet{peeples2011} found that the \mstar--$\zeta_\mathrm{wind}$ relation
follows the general shape of the direct method \mstar--$\zeta_\mathrm{wind}$
relation regardless of the input MZR (see their Figure 6).  However, the direct
method MZR requires more efficient metal ejection by winds than theoretical
strong line calibration MZRs (\citetalias{tremonti2004};
\citetalias{zaritsky1994}; \citetalias{kobulnicky2004};
\citetalias{mcgaugh1991}) at all stellar masses because of the lower
normalization of the direct method MZR.  We note that the yield is poorly
constrained, and a higher adopted yield requires more efficient outflows to
produce the observed MZR.

In contrast to the \citet{peeples2011} framework that assumed a zero scatter MZR
(and therefore does not account for variations in the SFR or gas fraction at a
fixed stellar mass), the \citet{finlator2008} model, based on cosmological
hydrodynamic simulations, treats the MZR as an equilibrium condition.  In their
model, galaxies are perturbed off the MZR by stochastic inflows but the star
formation triggered by the inflow of gas and the subsequent metal production
returns them to the mean MZR.  The rate at which galaxies re-equilibrate
following an episode of gas inflow sets the scatter in the MZR, which is
indirectly traced by the SFR-dependence of the \mstar--$Z$--SFR relation.

The observed SFR-dependence of the \mstar--$Z$--SFR relation differs according
to the strong line metallicity calibration used to construct the
\mstar--$Z$--SFR relation, as found by \citet{yates2012}.  Specifically, they
used metallicities estimated with the \citet{mannucci2010} method and
\citetalias{tremonti2004} method.  At low stellar masses, metallicity decreases
with increasing SFR for both \mstar--$Z$--SFR relations.  But at high stellar
masses (log[\mstar] $\gtrsim$ 10.5), the SFR-dependence of the
\citetalias{tremonti2004} \mstar--$Z$--SFR relation reverses, so that
metallicity increases with increasing SFR; however, the \citet{mannucci2010}
\mstar--$Z$--SFR relation collapses to a single sequence that is independent of
SFR.  \citet{yates2012} suggested that the SFR-dependence of the
\citet{mannucci2010} \mstar--$Z$--SFR relation at high stellar mass is obscured
by the N2 indicator (which was averaged with the metallicity estimated from
\rtt) used in the \citet{mannucci2010} metallicity calibration, which saturates
at high metallicity.

Unlike the \citet{mannucci2010} and \citetalias{tremonti2004} \mstar--$Z$--SFR
relations, the SFR-dependence of the direct method \mstar--$Z$--SFR relation
does not change dramatically with stellar mass.  There is little overlap between
the constant SFR tracks in the direct method \mstar--$Z$--SFR relation (Figure
\ref{fig:mzr_sfr}).  Furthermore, the SFR-dependence is strong ($\alpha$~=~0.66;
see Section \ref{sec:fmr}), so the scatter in the direct method MZR for
individual galaxies (if it could be measured) would be larger than the scatter
in the \citet{mannucci2010} and \citetalias{tremonti2004} MZRs. Within the
context of the \citet{finlator2008} model, this means that the direct method MZR
implies a longer timescale for galaxies to re-equilibrate than the
\citet{mannucci2010} and \citetalias{tremonti2004} MZRs.  We note that the
direct method \mstar--$Z$--SFR relation does not probe above log(\mstar)~=~10.5
because the auroral lines are undetected in this regime; however, this mass
scale is where the discrepancies between the \citet{mannucci2010} and
\citetalias{tremonti2004} metallicities are the largest---potentially due to a
break down of strong line calibrations at high metallicities (see Section
\ref{sec:alpha}).



\section{Summary}
\label{sec:summary}
We have measured \oiiis, \oiis, \niis, and \siis\ electron temperatures, direct
method gas-phase oxygen abundances, and direct method gas-phase nitrogen to
oxygen abundance ratios from stacked galaxy spectra.  We stacked the spectra of
$\sim$200,000 {\sc SDSS} star-forming galaxies in bins of (1) 0.1 dex in stellar
mass and (2) 0.1 dex in stellar mass and 0.5 dex in SFR.  The high SNR stacked
spectra enabled the detection of the temperature-sensitive auroral lines that
are essential for metallicity measurements with the direct method.  Auroral
lines are weak, especially in massive, metal-rich objects, but we detect \oiiia\
up to log(\mstar)~=~9.4 and \oiia\ up to log(\mstar)~=~10.5, which is generally
not feasible for spectra of individual galaxies.  We used the auroral line
fluxes to derive the \oiiis\ and \oiis\ electron temperatures, the O$^{++}$ and
O$^+$ ionic abundances, and the total oxygen abundances of the stacks.

We constructed the direct method mass--metallicity and \mstar--$Z$--SFR
relations across a wide range of stellar mass (log[\mstar]~=~7.4--10.5) and SFR
(log[SFR]~=~$-$1.0$\rightarrow$2.0).  The direct method MZR rises steeply
(O/H~$\propto$~\mstar$^{1/2}$) from log(\mstar)~=~7.4--8.9.  The direct method
MZR turns over at log(\mstar)~=~8.9, in contrast to strong line MZRs that
typically turn over at higher masses (log[\mstar]~$\sim$~10.5).  Above the
turnover, the direct method MZR approaches an asymptotic metallicity of
12~+~log(O/H)~=~8.8, which is consistent with empirical strong line calibration
MZRs but $\sim$0.3 dex lower than theoretical strong line calibration MZRs like
the \citet{tremonti2004} MZR.  Furthermore, we found that the SFR-dependence (as
measured by the value of $\alpha$ that minimizes the scatter at fixed
$\mu_\alpha \equiv \mathrm{log}(M_\star) - \alpha \mathrm{log(SFR)}$ in the
fundamental metallicity relation; see Equation \ref{eqn:fmr}) of the direct
method \mstar--$Z$--SFR relation is $\sim$2--3 times larger ($\alpha$~=~0.66)
than for strong line \mstar--$Z$--SFR relations ($\alpha \sim 0.12$--0.34).  Its
SFR-dependence is monotonic as a function of stellar mass, so constant SFR
tracks do not overlap, unlike strong line \mstar--$Z$--SFR relations.

We also showed that the direct method N/O relative abundance correlates strongly
with oxygen abundance and even more strongly with stellar mass.  N/O exhibits a
clear transition from primary to secondary nitrogen enrichment as a function of
oxygen abundance and stellar mass.

The slope, turnover, normalization, and SFR-dependence of the MZR act as
critical constraints on galaxy evolution models and are best measured by
methods that do not rely on strong line diagnostics, such as the direct method.
Future work should aim to construct a direct method MZR of individual galaxies
with high SNR optical spectra that enable the detection of auroral lines in
high mass and high metallicity objects.  Furthermore, metallicities based on
\textit{Herschel Space Observatory} \citep{pilbratt2010} measurements of the
far-IR fine-structure lines (Croxall et al.~in prep.) from the KINGFISH survey
\citep{kennicutt2011} will provide a valuable check on the absolute abundance
scale (see also \citealt{garnett2004a}), which is a major outstanding
uncertainty for galaxy evolution studies.  These types of investigations will
improve our understanding of the galaxy formation process, particularly the
cycling of baryons between galaxies and the IGM.


\acknowledgments

We gratefully acknowledge the referee, John~Moustakas, for his insightful
suggestions that improved this work.  We kindly thank Molly Peeples for her
detailed comments on a draft of this paper.  We are indebted to
Roberto~Cid~Fernandes, Kevin Croxall, Romeel~Dav{\'e}, Stacy~McGaugh,
Rick~Pogge, Rebecca~Stoll, and David~Weinberg for their feedback and
stimulating conversations.

We greatly appreciate the MPA/JHU group for making their catalog public.  We
thank the {\sc miles} team for making their SSPs publicly available.  This
research was partially based on data from the {\sc miles} project.  The {\sc
starlight} project is supported by the Brazilian agencies CNPq, CAPES and FAPESP
and by the France-Brazil CAPES/Cofecub program.  This research has made use of
NASA's Astrophysics Data System Service and the Cosmology Calculator by
\citet{wright2006}.

Funding for the SDSS and SDSS-II has been provided by the Alfred P.~Sloan
Foundation, the Participating Institutions, the National Science Foundation, the
U.S. Department of Energy, the National Aeronautics and Space Administration,
the Japanese Monbukagakusho, the Max Planck Society, and the Higher Education
Funding Council for England. The SDSS Web Site is http://www.sdss.org/.

The SDSS is managed by the Astrophysical Research Consortium for the
Participating Institutions. The Participating Institutions are the American
Museum of Natural History, Astrophysical Institute Potsdam, University of Basel,
University of Cambridge, Case Western Reserve University, University of Chicago,
Drexel University, Fermilab, the Institute for Advanced Study, the Japan
Participation Group, Johns Hopkins University, the Joint Institute for Nuclear
Astrophysics, the Kavli Institute for Particle Astrophysics and Cosmology, the
Korean Scientist Group, the Chinese Academy of Sciences (LAMOST), Los Alamos
National Laboratory, the Max-Planck-Institute for Astronomy (MPIA), the
Max-Planck-Institute for Astrophysics (MPA), New Mexico State University, Ohio
State University, University of Pittsburgh, University of Portsmouth, Princeton
University, the United States Naval Observatory, and the University of
Washington.




\end{document}